\documentclass[11pt]{article}
\newtheorem{theorem}{Theorem}

\newtheorem{proposition}{Proposition}

\usepackage[toc,page]{appendix}
\newenvironment{proof}{\noindent{\bf Proof:}}{\hfill\fbox{}\vspace*{1mm}}
\usepackage[usenames]{color}
\usepackage{amssymb}
\usepackage{graphicx}
\usepackage{subfigure}
\usepackage{float}
\usepackage{booktabs}

\RequirePackage[normalem]{ulem} 
\providecommand{\DIFdeltex}[1]{{\protect\color{red}\sout{#1}}}                      
\usepackage{xcolor}
\usepackage{float}
\usepackage{changebar}
\newif\ifdiff
\difftrue
\ifdiff
  \newcommand{\del}[1]{\DIFdeltex{#1}}

\else
  \newcommand{\del}[1]{}

\fi

\begin{document}
\title{\bf On Optimal Pricing Model for Multiple Dealers in a Competitive Market}
\author{
Wai-Ki Ching
\thanks{Corresponding author. Advanced Modeling and Applied Computing Laboratory,
Department of Mathematics, The University of Hong Kong, Pokfulam
Road, Hong Kong. E-mail: wching@hku.hk. }
\and Jia-Wen Gu
\thanks{Department of Mathematical Science,
University of Copenhagen, Denmark. E-mail: jwgu.hku@gmail.com.}
\and Qing-Qing Yang
\thanks{ Advanced Modeling and Applied Computing Laboratory,
Department of Mathematics, The University of Hong Kong, Pokfulam
Road, Hong Kong. E-mail: kerryyang920910@gmail.com.}
\and Tak-Kuen Siu
\thanks{ Department of Applied Finance and Actuarial Studies, Faculty of Business and Economics, Macquarie University, Sydney, NSW 2109, Australia.
Email: ktksiu2005@gmail.com}
}

\maketitle

\begin{abstract}
In this paper, the optimal pricing strategy in Avellande-Stoikov's \cite{Avellaneda08} for a monopolistic dealer
is extended to a general situation where multiple dealers are present in a competitive market.
The dealers' trading intensities, their optimal bid and ask prices and
therefore their spreads are derived
when the dealers are informed the severity of the competition.
The effects of various parameters
on the bid-ask quotes and profits of the dealers in a competitive market are also discussed.
This study gives some insights on the average spread, profit of the dealers in
a competitive trading environment.
\end{abstract}

\section{Introduction}

The role of a dealer in securities markets is to
stand ready immediately to trade fixed amounts
of securities at stated bid and ask prices.
An investor who would like to trade immediately
(a demander of immediacy)
can do so by placing a market order to trade at the best
available price:
the bid price if selling or the ask price if buying.
Liquidity provision was once performed only by dedicated broker-dealers,
known as market makers.
Market makers kept enough liquidity in their hands
so as to satisfy both supply and demand of any arriving traders.
In recent years, with the growth of electronic exchanges
such as Nasdaq's Inet, anyone wishing to submit limit orders
in the system can easily play the role of a dealer.
Agents who can post limit orders based on the availability of high frequency data, form a competitive trading environment.
In this paper, we focus on studying the optimal pricing strategies under multiple dealers' competition.

The pricing strategies of dealers have been studied extensively
in the micro-structure literature.
The two most often addressed sources of risk faced by dealers are:
(1) the inventory risk arising from uncertainty in the asset's value; and
(2) the asymmetric information risk arising from informed trades.
As noted by Demsetz (1968) \cite{Demsetz68},
while the limit orders are in a queue,
the dealer who places limit orders incurs both inventory and waiting costs.
Inventory costs arise from uncertainty about market prices of the securities that the dealer may hold in his portfolio while his limit orders are pending.
The waiting costs are the opportunity costs associated with the time between placing an order and its execution.
The concept of transaction costs was first proposed by
Demsetz \cite{Demsetz68}.
Copeland and Galai (1983) \cite{Copeland83}
pointed out in their paper that limit orders also
suffer from an informational disadvantage,
whereby they are picked off by better-informed investors.

Generally speaking, the probability of limit orders being
executed depends on the limit order price proximity
to the current market price.
Cohen et al. (1981) \cite{Cohen81},
called this phenomenon a ``gravitational pull'' of existing quotes.
Limit orders placed at current market quotes are likely to be executed,
whereas the probability of execution for aggressive limit orders
is close to zero.
They also pointed out that the probability of executing a limit order  decreases as a security's order arrival rate decreases.

Stoll (1978) \cite{Stoll78} developed an explicit and rigorous model for an individual dealer limited to the behavior of a single dealer making a market in a single stock.
They also discussed the implications of trading cost in different market organizations of dealers.
Ho and Stoll (1981) \cite{Ho81} extended Stoll's (1978) work
in the aspect of uncertainty (introduction of transactions uncertainty),
explicit treatment of a multi-period strategy for the dealer, and also the introduction of the demand side.
For the supply side, the research work was pioneered by Demsetz (1968) \cite{Demsetz68}, and additional theoretical contributions were made by Tinic (1972) \cite{Tinic72} and Stoll (1978) \cite{Stoll78}.

Garman (1976) \cite{Garman76} was the first to investigate the optimal market-making conditions through modeling temporary imbalances between buy and sell orders.
Ho and Stoll (1981) \cite{Ho81} adopted Garman's concept of
stochastic supply and demand for securities,
viewing both the demand to sell and buy securities
as demand for dealer services and assuming a linear demand relation
for dealer sales and purchases.
They took the ``true'' price of the stock to be exogenously determined by an information set, and assumed that the dealer' prices were relative to
the ``true'' price.
The key concern of Ho and Stoll is about the risk that the dealer faces,
and how this affected dealer's willingness to provide his services.
In another paper by Ho and Stoll (1980) \cite{Ho80},
the problem of dealers under competition was analyzed and the
bid and ask prices were shown to be related to the reservation
(or indifference) prices of the agents.

Based on the idea and work of Ho and Stoll (1981) \cite{Ho81},
Avellaneda and Stoikov (2008) \cite{Avellaneda08}
enhanced Garman's model to a quantitative market-making limit order book strategy that generates persistent positive returns;
indeed the economic setting of the problem is similar.
The main differences are the nature of ``true'' price of the asset,
the explicit utility function of the agent and the trading intensity
under the laws governing the micro-structure of financial markets.
In \cite{Avellaneda08},
the ``true'' price was given by the market mid-price.
In order to model the arrival rate of buy and sell orders
that will reach the agent, they drew on recent results in econophysics
\footnote{Econophysics is an inter-disciplinary research area,
which applies theories and methods in Physics to study Economics problems.},
Potters and Bouchaud (2003) \cite{Potters2003},
and gave an exponential arrival rate of the market orders.
The approach adopted is to combine the utility framework of the
Ho and Stoll (1981) \cite{Ho81} approach with the micro-structure of actual limit order book as described in the econophysics literature.
The strategy, focusing on the effects of inventory risk, outperforms the ``best-bid-best-ask'' market-making strategy where the trader posts limit orders at the best bid and ask available on the market.

More recently, Cartea and Jaimungal \cite{CJ} used a similar model to introduce risk measures for high-frequency trading.
They used a model inspired by Avellaneda-Stoikov  \cite{Avellaneda08} in which the mid-price is modeled by a Hidden Markov Model (HMM).
Guilbaud and Pham \cite{GP} also used a model inspired by the Avellaneda-Stoikov framework but including market orders and limit orders the best
(and next to the best) bid and ask together with stochastic spreads.
Gu{\'e}ant et al. \cite{GLF} provided some simple and easy-to-compute expressions for the optimal quotes when the trader is willing to liquidate a portfolio.
Some other recent works include Alavi Fard (2014) \cite{F.A2014} and
Song et al. (2012) \cite{N.Song2012} for example.
It seems that in the literature more attention has been given to
trading strategies in a single dealer's framework while
relatively little attention has been paid to a multiple-dealer case.
We are dedicated to the latter case.

In this paper, the dealers' optimal bid and ask quotes in the multiple dealers case are determined.
The optimal pricing strategy accounts for two key factors: the inventory level of the the dealer and the competition severity.
Our paper contributes to the high frequency trading literature in various aspects.
Firstly, we derive the optimal bid and ask prices for each dealer when the dealers are informed the severity of the competition (for example,how many active dealers are in the market). Each dealer quotes his optimal prices considering the market information and his own inventory level to maximize his final profit at terminal time $T$.
Secondly, we compare quoting prices with those obtained in \cite{Avellaneda08} to shed some lights on trading competition in the market.
Thirdly, we also conduct comparison of the profit generated by dealers in competitive markets with that in single dealer markets. This comparison may hopefully enhance our understanding on how high frequency trading dealers gain profits by providing stock liquidity.

The remainder of the paper is structured as follows.
In Section 2, to give readers the background of research,
we review two models, one by Ho and Stoll in \cite{Ho81}
and the other by Avellaneda and Stoilov in \cite{Avellaneda08}.
In Section 3, we present the extended model
and results in \cite{Avellaneda08} for a market consisting of $N$ dealers.
The limit order book intensities for the dealers are derived.
In Section 4,
we consider two situations, inactive dealers and active dealers,
for the multiple dealer problem.
In the case of dealers in markets under competition,
we need to make use of an approximation method and the
principle of Dynamic Programming (DP), i.e.,
the solution methods combine backward induction
with a forward simulation  of states.
In Section 5, numerical results are given and compared for one-dealer and
multiple-dealer situations.
Finally, concluding remarks are given in Section 6
to discuss further research issues.

\section{A Review of Two Models}

In this section, we describe the background of this research by
reviewing two related models, one by Ho and Stoll
in \cite{Ho81} and the other by  Avellaneda and Stoilov in  \cite{Avellaneda08}.
The developments below follow those in \cite{Avellaneda08,Ho81}.

\subsection{Ho-Stoll Model}

The model proposed by Ho and Stoll (1981) \cite{Ho81} imposed the
following assumptions.

\begin{enumerate}
\item[(i)] Transactions are assumed to evolve over time according to Poisson jump processes as in Garman (1976) \cite{Garman76}.
Two Poisson processes are used, one for purchases by the dealer and
the other for sales by the dealer, as follows:
$$
dq_a={\cal X}_{\{\mbox{an arrival of a market buy order}\}} Q\lambda_a dt
$$
and
$$
dq_b={\cal X}_{\{\mbox{an arrival of a market sale order}\}}Q\lambda_bdt
$$
where ${\cal X}_{E}$ is the indicator function of the event $E$,
$Q$ is the market order size, and
$\lambda_a$ and $\lambda_b$ are the {\it intensities} of the transactions.
Here $d q_a$ and $d q_b$ are the increments of the number of
market buy orders and the number of market sale orders, respectively.

\item[(ii)] The dealer determines a price of immediacy, $b$, should a market sale order arrive and a price, $a$, should a market buy order arrive.
The dealer does not directly quote $b$ and $a$,
instead, he quotes his bid and ask prices, respectively, as follows:
$$
p_b=p-b \quad {\rm and} \quad p_a=p+a.
$$
Here $p$ is the dealer's opinion
of the true price of the stock at the time he sets
the bid-ask quotation and this price is supposed to be a given constant.

\item[(iii)] The intensities,
  $\lambda_a$ and $\lambda_b$, depend on the dealer's
selling fee and buying fee, respectively.

\item[(iv)] In addition to uncertainty about the timing of subsequent transactions,
the dealer faces uncertainty about the return on his existing portfolio. Consequently
$$
\left\{
\begin{array}{lll}
dF &=& rFdt-(p-b)dq_b+(p+a)dq_a \\
dI &=& r_IIdt+pdq_b-pdq_a+IdZ_I\\
dY &=&r_YYdt+YdZ_Y.
\end{array}
\right.
$$
Here $F,I,Y$ are the balances of the cash account, inventory account, and base wealth,  respectively.
Here $r_I$, $r_Y$  represent the mean return of inventory account and base wealth per unit time, respectively.
And $r$ is the constant continuously compounded risk-free rate.
Here $Z_I$ and $Z_Y$ are Wiener processes with mean zero and constant variance rates, $\sigma_I^2$ and $\sigma_Y^2$, respectively.
\end{enumerate}

The objective of the dealer is to maximize the expected utility of his total wealth, $E_t[U(W_T)]$, at the terminal time $T$, where
$$
W_T=F_T+I_T+Y_T.
$$
Notice that $E_t[U(W_T)]$ is the conditional expectation of $U(W_T)$ given information up to time $t$.  Numerous transactions and price changes occur between $t$ and $T$.

\subsection{Avellaneda-Stoikov Model}

Avellaneda and Stoikov (2008) \cite{Avellaneda08}
modified the model of Ho and Stoll (1981) \cite{Ho81}
in some aspects.

\begin{enumerate}
\item[(i)] Assume that the money market pays no interest, and the mid-market price,
or mid-price, of the stock evolves over time according to the following zero-drift diffusion process:
\begin{equation}\label{eq1}
dS_u=\sigma dW_u
\end{equation}
where the initial value $S_t=s$, $\{W_u\}_{t\le u \le T}$ is a standard one-dimensional Brownian motion and $\sigma$ is a constant, i.e., a constant
volatility model.

\item[(ii)] The agent's objective is to maximize the expected exponential utility of his portfolio at the terminal time $T$.
The exponential utility is given by
\begin{equation}\label{eq2}
u(w) = -\exp(-\gamma w) 
\end{equation}
where $\gamma$ is the {\it risk-aversion} parameter.

\item[(iii)] The Poisson intensity at which the agent's orders are executed is supposed to be exponential.
In the symmetric case, exponential arrival rates are assumed to take the
following form:
\begin{equation}\label{eq3}
\lambda(\delta)=Ae^{-k \delta}
\end{equation}

\item[(iv)] The reservation bid and ask prices  $r^b(s,q,t)$ and $r^a(s,q,t)$, which can be interpreted as the indifference prices for buying and selling, respectively, 
are introduced and they satisfy
\begin{equation}\label{eq4}
\left\{
\begin{array}{l}
v(x-r^b(s,q,t),s,q+1,t)=v(x,s,q,t)\\
v(x+r^a(s,q,t),s,q-1,t)=v(x,s,q,t)
\end{array}
\right.
\end{equation}
where $v(x,s,q,t)=E_t[U(W_T)]$, $x$ is the initial wealth at time $t$ and $q$ is the initial inventory level at time $t$. 
\end{enumerate}

In their model, it is assumed that there is only one
monopolistic dealer in the trading system.
The dealer buys or sells one stock in the market.
The dealer quotes the bid price $p^b$ and the ask price $p^a$,
and is committed to, respectively,
buy and sell one share of stock at these prices.
The wealth in cash $X_t$ jumps whenever there is a buy order
or sell order and it is governed by
\begin{equation}\label{eq5}
d X_t =p^a d N_t^a - p^b d N_t^b. 
\end{equation}
Here $N_t^b$ is the amount of stocks bought by the dealer and $N_t^a$
is the amount of stocks sold.
They are assumed to follow Poisson processes with intensities
$\lambda^b$ and $\lambda^a$, respectively.
The number of units of the stock or the
inventory level held by the dealer is then governed by
\begin{equation}\label{eq6}
dq_t = dN_t^b-dN_t^a. 
\end{equation}
The objective of the dealer who can set limit orders is
\begin{equation}\label{eq7}
u(s,x,q,t)= \displaystyle\max_{\delta^a, \delta^b} E_t \left[-\exp(-\gamma(X_T + q_T S_T))\right]
\end{equation}
where $\delta^a=p^a-s$, $\delta^b = s-p^b$
and the dealer holds $q$ stocks at time $t$.
Note that $\delta^a$ and $\delta^b$ are the prices of immediacy for selling and buying, respectively, from the dealer's perspective.
For the case of an inactive trader, the choice variables $\delta^a$ and $\delta^b$ are absent. According to Avellaneda and Stoikov (2008) \cite{Avellaneda08},
the value function for the inactive dealer,
who holds an inventory of $q$ stocks until the terminal time $T$,
can be written as follows:
\begin{equation}\label{eq8}
u(x,s,q,t) = -\exp(\gamma x) \exp (-\gamma qs)
\exp\left(\frac{\gamma^2q^2\sigma^2(T-t)}{2}\right).
\end{equation}
The definition of reservation or indifference price was introduced in
\cite{Avellaneda08} and we shall give it in Eq. (\ref{eq11}).
The dealer's reservation bid and ask prices $r^b$ and $r^a$ are given, respectively, by
\begin{equation}\label{eq9}
\left\{
\begin{array}{l}
u(x-r^b(s,q,t),s,q+1,t) = u(x,s,q,t)\\
u(x+r^a(s,q,t),s,q-1,t) = u(x,s,q,t)
\end{array}
\right.
\end{equation}
i.e.,
\begin{equation}\label{eq10}
\left\{
\begin{array}{l}
r^b (s,q,t) = s +(-1-2q)\frac{\gamma\sigma^2(T-t)}{2}\\
r^a (s,q,t) = s +(1-2q)\frac{\gamma\sigma^2(T-t)}{2}.
\end{array}
\right.
\end{equation}
Hence the average of the above two prices,
say the {\it reservation price} or the {\it indifference price}, is given by
\begin{equation}\label{eq11}
r(s,q,t) = s - \gamma q \sigma^2 (T-t).
\end{equation}
To consider the case of active dealers,
who will make decisions to buy or sell before the terminal time $T$,
in \cite{Avellaneda08}, the authors derived the following HJB equation
(see for instance \cite{Avellaneda08}, Section 3):
\begin{equation}\label{eq12}
\left\{
\begin{array}{l}
\displaystyle \frac{\partial u}{\partial t} + \frac{1}{2}\sigma^2 \frac{\partial^2 u}{\partial s^2} + \max_{\delta^b}
\lambda^b \left[u(s, x-s+\delta^b, q+1,t)-u(s, x,q,t)\right]\\
 +\displaystyle  \max_{\delta^a} \lambda^a \left[u(s, x + s +\delta^a, q-1,t)
 -u(s, x,q,t)\right]=0\\
u (s,x,q,T)=-\exp(-\gamma (x +qs)).\\
\end{array}
\right.
\end{equation}
To solve the HJB equation, in \cite{Avellaneda08}, the authors
consider the simplest case by  assuming
that the Poisson intensities take the following form
(c.f. Eq. (\ref{eq3})):
\begin{equation}\label{eq13}
\lambda^b (\delta) = \lambda^a (\delta) = A e^{-k\delta}.
\end{equation}
Then the following trial solution was adopted:
\begin{equation}\label{eq14}
u(s,x,q,t) = -\exp(-\gamma x)\exp ( -\gamma\theta(s,q,t))
\end{equation}
where $\theta(s,q,t)$ is approximated up to the second-order
of a Taylor expansion  about the inventory variable $q$:
\begin{equation}\label{eq15}
\theta(s,q,t) = \theta^{(0)}(s,t) + \theta^{(1)}(s,t)q
+ \theta^{(2)}(s,t)q^2 + \ldots + .
\end{equation}
When the inventory level is $q$, the reservation bid price of the stock
and the reservation ask price of the stock are given, respectively, by
\begin{equation}\label{eq16}
\left\{
\begin{array}{l}
r^b(s,q,t) = \theta(s,q+1,t) - \theta (s,q,t)\\
r^a(s,q,t) = \theta(s,q,t) -\theta(s,q-1,t).
\end{array}
\right.
\end{equation}
Substituting $\theta$ into Eq. (\ref{eq12}) yields
\begin{eqnarray}\label{eq17}
\left\{
\begin{array}{l}
\theta_t + \frac{1}{2}\sigma^2\theta_{ss} - \frac{1}{2}\sigma^2 \gamma \theta_s^2
+\displaystyle \max_{\delta^b}\left(  \frac{\lambda^b(\delta^b)}{\gamma}(1-e^{\gamma(s-\delta^b-r^b)} \right)\\
+\displaystyle \max_{\delta^a}\left(  \frac{\lambda^a(\delta^a)}{\gamma}(1-e^{-\gamma(s+\delta^a-r^a)} \right)=0\\
\theta(s,q,T)=qs.
\end{array}
\right.
\end{eqnarray}
Using the first-order optimality condition,
the problem can be transformed into the following one:
\begin{equation}\label{eq18}
\left\{
\begin{array}{l}
\theta_t + \frac{1}{2}\sigma^2 \theta_{ss}-\frac{1}{2}\sigma^2\gamma \theta_s^2
+\frac{A}{k+\gamma} \left( e^{-k\delta^a} + e^{-k\delta^b}\right)=0\\
\theta(s,q,T)=qs.
\end{array}
\right.
\end{equation}
In \cite{Avellaneda08}, the authors consider
an asymptotic expansion of $\theta$ about $q$,
and higher order terms are assumed to be small enough to be negligible.
By considering the coefficients of $q$ and $q^2$,
the following results are obtained:
\begin{equation}\label{eq19}
r(s,t)= s- q\gamma \sigma^2(T-t)
\end{equation}
which matches with Eq. (\ref{eq11}) and the bid-ask spread is then given by
\begin{equation}\label{eq20}
\delta^b +\delta^a=\gamma\sigma^2(T-t)
+\frac{2}{\gamma}\ln \left(1 + \frac{\gamma}{k}\right).
\end{equation}

\section{The Limit Order Book Rates}

In this section, we consider the situation of multiple dealers in a competitive market.
We focus on the discussion of  dealers' trading rates in the market.

In \cite{Avellaneda08}, for only one dealer in the market,
the arrival rates of buy and sell orders that will reach the
dealer follows a Poisson process with a common exponential arrival
rate in Eq. (\ref{eq13}) which depends on dealer's quotes.
In the case of multiple dealers market under competition, we assume that the overall frequency of market orders will depend on the quotes of all dealers in the market.
Suppose that there are $N$ dealers in the market and that dealers have
impacts on the arrival of market orders.
Under certain assumptions, we shall show that the market orders
follow a Poisson process with common exponential arrival rate,
$$
\lambda^a(\delta_1,\cdots,\delta_N)=\lambda^b(\delta_1,\cdots,\delta_N)=A e^{-k(\beta_1\delta_1+\cdots+\beta_N\delta_N)}.
$$
Here $\beta_i$ reflects the influence (e.g. competitiveness) of Dealer $i$
on the overall frequency of  market orders.
In a number of studies \cite{Avellaneda08,Bouchaud02,Gabaix06,Gopikrishnan00,Maslow01},
it has been shown that the distances $\delta_i^a$, $\delta_i^b$ (c.f. Eq. (\ref{eq7})) and the current shape of the limit order book determine the priority of execution when large market orders get executed.
For example, when a large market order to buy $Q$ stocks arrives,
the $Q$ limit orders with the lowest ask prices will be automatically executed.
Let $p^Q$ be the price of the highest limit order executed in this trade, we define $\Delta p=p^Q-s$ to be the temporary market impact of the trade of size $Q$.
If the agent's limit order is within the range of this market order, i.e. $\delta_i^a<\Delta p$, his limit order will be executed.
To quantify dealers' trading intensity, other than the overall frequency of market orders, we need to know the distribution of market orders' size
and the temporary impact of a large market order.
We have the following proposition.

\begin{proposition}\label{Prop1}
Suppose that dealers' impacts on the overall frequency of market orders, i.e., $\lambda^a(\delta_1^a,\ldots,\delta_N^a)$ and $\lambda^b(\delta_1^b,\ldots,\delta_N^b)$ are ``separable'' and have an ``identical functional form'', i.e.,
(we take $\lambda^a$ as an example), taking the following form:
$$\lambda^a(\delta_1^a,\cdots,\delta_N^a)=
f_1(\delta_1^a)f_2(\delta_2^a)\cdots f_N(\delta_N^a).
$$
and $f_i(\delta_i^a)=f(\delta_i^a)^{\beta_i}$.
Here $\beta_i$ describes Dealer $i$'s impact on market orders' overall frequency.
Then 
$$
\lambda^a(\delta_1^a,\ldots,\delta_N^a)
=Ae^{-k(\beta_1 \delta_1^a +\cdots +\beta_N\delta_N^a)}.
$$
Furthermore, if the distribution of the size of market orders $Q$ obeys a
``power law'', \cite{Gabaix06,Gopikrishnan00,Maslow01},
 i.e., $f^Q(x)\varpropto x^{-1-\alpha}$
and the market impact follows a ``$\log$ law'' \cite{Bouchaud02},
i.e., $\Delta p \varpropto \ln(Q)$, then we have
$$
\lambda_i^a=A e^{-k(\beta_1 \delta_1^a +\cdots +\beta_N\delta_N^a)}e^{-(1-\frac{1}{N})\beta_i\delta_i^a}.
$$
\end{proposition}

The proof of the proposition can be found in Appendix A.

\section{The Multiple-Dealer Problem}

In this section, we discuss the situation of multiple dealers buying
and selling stocks in a competitive market.
In particular, we consider two situations: inactive and active dealers.
The underlying problem is a state-feedback control problem.
Both inactive dealer's ``frozen'' strategy and
active dealer's optimal quoting strategy are
discussed in this section.
For the active dealer's situation, each dealer
follows a multi-period strategy that maximizes
his objective function taking into account not only his own possible
future actions but also those of his competitors.
This is much more complex than the single dealer case.
Discrete and continuous models for active dealers are discussed.
Furthermore, a comparison between the performances
of active dealers and inactive dealers is made.

Our objective is to study the trading strategies of different dealers
in a competitive market, where each dealer has his own reservation value
of the stock based on his inventory position.
The dealers wish to buy or sell stocks in the market,
and the mid-price is assumed to be governed by
the following stochastic differential equation (c.f. Eq. (\ref{eq1})):
$dS_u = \sigma d W_u$ with initial value $S_t = s$.
Here $\{W_u\}$ is a standard Brownian motion
and $\sigma$ is a positive constant.
Each dealer $i$, $(i=1,2,\ldots,N)$, quotes his bid price $p_i^b(u)$
and ask price $p_i^a(u)$, and is committed to, respectively,
buying and selling one share of the stock at these prices at time $u$.
Hence, the wealth of Dealer $i$ in cash $X_i(u)$ jumps
whenever there is a buy or sell order executed.
\begin{equation}\label{eq21}
d X_i(u) =p_i^a(u) d N_i^a(u) - p_i^b(u) d N_i^b(u)
\end{equation}
where $N_i^b(u)$
is the amount of stocks bought and $N_i^a(u)$
is the amount of stocks sold by Dealer $i$ up to time $u$.
They are supposed to follow Poisson processes
with intensities, $\lambda_i^b$ and $\lambda_i^a$, respectively.
In view of the results in Proposition 1,
the Poisson intensities take the form:
\begin{equation}\label{eq22}
\left\{
\begin{array}{l}
\lambda_i^a=\displaystyle A e^{-k(\beta_1 \delta_1^a +\cdots +\beta_N\delta_N^a)}e^{-\beta_i(1-\frac{1}{N})\delta_i^a}\\
\lambda_i^b=\displaystyle A e^{-k(\beta_1 \delta_1^b +\cdots +\beta_N\delta_N^b)}e^{-\beta_i(1-\frac{1}{N})\delta_i^b}.
\end{array}
\right.
\end{equation}
The number of stocks is governed by
\begin{equation}\label{eq23}
dq_i(u) = dN_i^b(u)-dN_i^a(u).
\end{equation}
Let
\begin{equation}\label{eq24}
\delta_i^a(u)=p_i^a(u)-S_u \quad  {\rm and } \quad \delta_i^b(u)=S_u-p_i^b(u)
\end{equation}
then we have
\begin{equation}\label{eq25}
\begin{array}{lll}
d(X_i(u)+q_i(u)S_u)&=&dX_i(u)+S_udq_i(u)+q_i(u)dS_u\\
&=&p_i^a(u)dN_i^a(u)-p_i^b(u)dN_i^b(u)+S_u(dN_i^b(u)-dN_i^a(u))+q_i(u)\sigma dW_u\\
&=&\delta_i^a(u)dN_i^a(u)+\delta_i^b(u)dN_i^b(u)+q_i(u)\sigma dW_u.\\
\end{array}
\end{equation}

\subsection{Inactive Dealer}

We first consider an inactive trader, Dealer $i$,
who does not have any limit orders in the market and
simply holds an inventory of $q_i$ stocks until the terminal time $T$,
which is a special case of the feedback control problem in which $(\delta_i^a,\delta_i^b)=(\infty,\infty)$. Following \cite{Avellaneda08},
it is not difficult to show that
\begin{equation}\label{eq26}
u_i(s,x_i,q_i,t)=-\exp(-\gamma_ix_i)\exp(-\gamma_iq_is)\exp \left(\frac{\gamma_i^2q_i^2\sigma^2(T-t)}{2}\right)
\end{equation}
which is the same as the value function calculated in the monopolistic market,
showing us directly its dependence on the market parameters.
The reservation bid and ask prices are given implicitly by the relations
\begin{equation}\label{eq27}
\left\{
\begin{array}{l}
v_i(s,x_i-r_i^b(s,q_i,t),q_i+1,t)=v_i(s,x_i,q_i,t)\\
v_i(s,x_i+r_i^a(s,q_i,t),q_i-1,t)=v_i(s,x_i,q_i,t)
\end{array}
\right.
\end{equation}
which means that the agent is indifferent between keeping
inactive and buying one stock at the reservation bid price $r_i^b$
(or, selling one stock at the reservation ask price $r_i^a$).
It is straightforward to calculate that
\begin{equation}\label{eq28}
\left\{
\begin{array}{l}
r_i^b(s,q_i,t)=s-(1+2q_i)\frac{\gamma_i\sigma^2(T-t)}{2}\\
r_i^a(s,q_i,t)=s+(1-2q_i)\frac{\gamma_i\sigma^2(T-t)}{2}
\end{array}
\right.
\end{equation}
and the reservation (or difference) price is given by
\begin{equation}\label{eq29}
r_i(s,q_i,t)=\frac{r_i^a(s,q_i,t)+r_i^b(s,q_i,t)}{2} =s-q_i\gamma_i\sigma^2(T-t).
\end{equation}

\subsection{The Active Dealers}

In general, it may not be easy to determine the optimal quoting
strategies for dealers in a competitive market.
In the market, each dealer's action depends not only on his own but also his competitor's characteristics.
They all need to solve a relatively complex Dynamic Programming (DP)
problem than the one encountered in the single dealer case.
In this section, we develop a feasible quoting policy  using
a linear approximation method and the principle of DP.
We first discuss a discrete model and
then give a recursive formula for the bid and ask quotes.
We then extend the discrete model to a continuous one.
By directly using a linear approximation and the DP principle to solve the optimal
control problem, one can obtain an optimal quoting strategy for dealers in the continuous competition model.

\subsubsection{The One-period Model}

Suppose there are $N$ dealers in the market,
namely Dealer 1, Dealer 2, $\ldots$, Dealer $N$.
In the one period case, we assume that dealers may only trade in the last trading session, $(t_{n-1},t_n)$ and trades happen immediately after time $t_{n-1}$. Dealers choose their bid and ask quotes at the beginning of the trading session, $t_{n-1}$, defined through the controls $(\delta_{n-1}^{i,b},\delta_{n-1}^{i,a})_{i=1}^N$.
These quotes influence the arrival rates of market orders
over the time interval $(t_{n-1,},t_n)$.
By Eq. (\ref{eq22}), the arrival rates take the following forms:
\begin{equation}\label{eq30}
\left\{
\begin{array}{lll}
\lambda_{n-1}^{i,a}&=&\displaystyle
Ae^{-k(\beta_1\delta_{n-1}^{1,a}+\cdots+\beta_N\delta_{n-1}^{N,a})}e^{-(1-\frac{1}{N})\beta_i\delta_{n-1}^{i,a}}\\
\lambda_{n-1}^{i,b}&=&\displaystyle Ae^{-k(\beta_1\delta_{n-1}^{1,b}+\cdots+\beta_N\delta_{n-1}^{N,b})}e^{-(1-\frac{1}{N})\beta_i\delta_{n-1}^{i,b}}.
\end{array}
\right.
\end{equation}
For any dealer in this competitive market, the objective is to determine the optimal bid and ask quotes to maximize his own expected utility function:
\begin{equation}\label{eq31}
V^i(s_{n-1},x_{n-1}^i,\gamma_1,\cdots,\gamma_N,q_{n-1}^1,\cdots,q_{n-1}^N,t_{n-1})=\displaystyle \max_{\delta_{n-1}^{i,a},\delta_{n-1}^{i,b}}\left\{E\left[-\exp( -\gamma_i\left(X_T^i+q_T^iS_T\right))|\mathcal{F}_{n-1} \right]\right\}
\end{equation}
which is a stochastic feedback control problem.
For any dealer, he can only determines his optimal bid and ask quotes $\delta_{n-1}^{i,b}$ and $\delta_{n-1}^{i,a}$.
However, the stochastic feedback problem is related to optimal bid and ask quotes  $\delta_{n-1}^{1,b},\cdots,\delta_{n-1}^{N,b}$ and $\delta_{n-1}^{1,a},\cdots,\delta_{n-1}^{N,a}$ of all dealers.
Thus it is also a problem of game competition,
especially, it is a simultaneous game problem.
Suppose all dealers achieve the Nash equilibrium in this game problem,
then the results in the proposition below follow.
\begin{proposition}
The optimal quoting policy in the one period case is
\begin{equation}\label{eq32}
\left\{
\begin{array}{lll}
\delta_{n-1}^{i,a}&=&\displaystyle\frac{1}{\gamma_i}\ln \left( 1+\frac{\gamma_i}{(k+1-\frac{1}{N})\beta_i}\right)+
\frac{\gamma_i \sigma^2 (T-t_{n-1})}{2}(-2q_{n-1}^i+1)\\
\delta_{n-1}^{i,b}&=&\displaystyle\frac{1}{\gamma_i}\ln \left( 1+\frac{\gamma_i}{(k+1-\frac{1}{N})\beta_i}\right)+
\frac{\gamma_i \sigma^2 (T-t_{n-1})}{2}(2q_{n-1}^i+1)\\
\end{array}
\right.
\end{equation}
and Dealer $i$'s utility is given by
\begin{equation}\label{eq33}
\begin{array}{lll}
&&V^i(s_{n-1}, x_{n-1}^i,
\gamma_{1},\cdots,\gamma_N,q_{n-1}^{1},\cdots,q_{n-1}^{N}, t_{n-1})\\
&=&\displaystyle -\exp\left(-\gamma_i(x_{n-1}^i+q_{n-1}^is_{n-1})\right)
\exp \left(\frac{\gamma_i^2\sigma^2(q_{n-1}^i)^2(T-t_{n-1})}{2}\right)
\\ 
&&\left[1-\frac{\gamma_i \Delta t_{n-1}}{(k+1-\frac{1}{N})\beta_i+\gamma_i}\left(\lambda_{n-1}^{i,a}+\lambda_{n-1}^{i,b}\right)\right]
\end{array}
\end{equation}
where $\Delta t_{n-1}=t_n-t_{n-1}$.
\end{proposition}

The proof can be found in Appendix B.

We remark that
\begin{description}
\item [(i)] Only in the one-period case, a dealer's bid and ask quotes are independent of his competitors.
However, even in the one-period case,
the value function of each dealer is not independent of the inventory position and other parameters, such as risk aversion of the competing dealers.

\item [(ii)] Dealer $i$'s bid-ask spread is given by
\begin{equation}\label{eq35}
\delta_{n-1}^{i,b}+\delta_{n-1}^{i,a}
=\displaystyle\frac{2}{\gamma_i} \ln \left( 1+\frac{\gamma_i}{(k+1-\frac{1}{N})\beta_i}\right)+
\gamma_i \sigma^2 (T-t_{n-1})
\end{equation}
which is independent of the inventory.
After taking a first-order approximation of the order arrival term, we have
\begin{equation}\label{eq36}
\lambda_{n-1}^{i,a}+\lambda_{n-1}^{i,b}=A\left[2-(k+1-\frac{1}{N})\beta_i(\delta_{n-1}^{i,a}+\delta_{n-1}^{i,b})-k\displaystyle\sum_{j\ne i}\beta_j(\delta_{n-1}^{j,a}+\delta_{n-1}^{j,b})+\cdots +\right]
\end{equation}
where the linear term does not depend on the inventory variables.
Therefore, if we substitute Eq. (\ref{eq36}) into Eq. (\ref{eq33}),
we arrive at the conclusion that Dealer $i$'s utility depends
only on his own inventory $q_{n-1}^i$.
We define this approximation as
$$
f^i(s_{n-1},x_{n-1}^i,q_{n-1}^i,\gamma_1,\cdots,\gamma_N,t_{n-1})
$$
which equals
\begin{equation}\label{eq37}
\displaystyle -\exp\left(-\gamma_i(x_{n-1}^i+q_{n-1}^is_{n-1})\right)
\exp \left(\frac{\gamma_i^2\sigma^2(q_{n-1}^i)^2(T-t_{n-1})}{2}\right)h_{n-1}^i
\end{equation}
where
\begin{equation}\label{eq38}
\begin{array}{lll}
h_{n-1}^i&=&\displaystyle 1-\frac{A\gamma_i\Delta t_{n-1}}{(k+1-\frac{1}{N})\beta_i+\gamma_i}
\Big\{2-(k+1-\frac{1}{N})\beta_i\Big[\frac{2}{\gamma_i}
\ln\left(1+\frac{\gamma_i}{(k+1-\frac{1}{N})\beta_i}\right)\\
&+&\displaystyle \gamma_i\sigma^2(T-t_{n-1})\Big]-k\sum_{j\ne i}\beta_j\Big[\frac{2}{\gamma_i}\ln\left(1+\frac{\gamma_j}{(k+1-\frac{1}{N})\beta_j}\right) +\gamma_j\sigma^2(T-t_{n-1})\Big]
\Big\}.
\end{array}
\end{equation}

\item [(iii)] Set
$$
f^i(s_{n-1},x_{n-1}^i,q_{n-1}^i,\gamma_1,\cdots,\gamma_N,t_{n-1})=-\exp\left(-\gamma_i(x_{n-1}^i+q_{n-1}^is_{n-1})\right)
g_{n-1}^i(q_{n-1}^i,t_{n-1})
$$
where
$$
g_{n-1}^i(q_{n-1}^i,t_{n-1})=\exp\left(\frac{\gamma_i^2\sigma^2(q_{n-1}^i)^2(T-t_{n-1})}{2}\right)h_{n-1}^i
$$
is independent of stock price $s_{n-1}$ and cash wealth $x_{n-1}^i$.

\item[(iv)] We define the market bid and ask quotes,
$$
\delta_{n-1}^b=\min\{\delta_{n-1}^{i,b},i=1,2,\ldots,N\}
\quad {\rm and} \quad
\delta_{n-1}^a=\min\{ \delta_{n-1}^{i,a},i=1,2,\ldots,N\}
$$
and the market bid-ask spread,
$$
s_{n-1}=\delta_{n-1}^b+\delta_{n-1}^a
$$
which depends on dealers' inventories.
We remark that Dealer $i$'s bid-ask spread is always positive,
however, the market bid-ask spread can be negative.

\item [(v)] Notice that
$$
\begin{array}{lll}
&&
V^i(s_{n-1},x_{n-1}^i,\gamma_1,\cdots,\gamma_N,
q_{n-1}^1,\cdots,q_{n-1}^N,t_{n-1})\\
&=&\displaystyle -\exp\left(-\gamma_i(x_{n-1}^i+q_{n-1}^is_{n-1})\right)
\exp \left(\frac{\gamma_i^2\sigma^2(q_{n-1}^i)^2(T-t_{n-1})}{2}\right)\\
&&\left[1-\frac{\gamma_i \Delta t_{n-1}}{(k+1-\frac{1}{N})\beta_i+\gamma_i}\left(\lambda_{n-1}^{i,a}+\lambda_{n-1}^{i,b}\right)\right]\\
&>& \displaystyle -\exp\left(-\gamma_i(x_{n-1}^i+q_{n-1}^is_{n-1})\right)
\exp \left(\frac{\gamma_i^2\sigma^2(q_{n-1}^i)^2(T-t_{n-1})}{2}\right)\\
\end{array}
$$
which means that active dealers will always have advantage over the
inactive dealers.
\end{description}

In the next section, we shall employ this linear approximation technique
to analyze the dynamics of dealer markets in the multi-period case.

\subsubsection{The Two-period Model}

Assume that Dealer $i$ may only trade in the intervals
$(t_{n-2},t_{n-1})$ and $(t_{n-1},t_n)$.
The dealer chooses bid and ask quotes at time $t_{n-2}$ and $t_{n-1}$ with
the controls $\delta_{n-2}^{i,b}, \delta_{n-2}^{i,a}, \delta_{n-1}^{i,b}$ and $\delta_{n-1}^{i,a}$, and trades happen immediately after time $t_{n-2}$ and $t_{n-1}$.
Adopting the above linear approximation,
one can establish the following proposition.

\begin{proposition}
In two period model, dealers' optimal bid and ask quotes are given by
\begin{equation}\label{eq39}
\left\{
\begin{array}{lll}
\delta_{n-2}^{i,b}&=&\displaystyle\frac{1}{\gamma_i}\ln \left( 1+\frac{\gamma_i}{(k+1-\frac{1}{N})\beta_i}\right)+
\frac{\gamma_i \sigma^2 (T-t_{n-2})}{2}(2q_{n-2}^i+1)\\
\delta_{n-2}^{i,a}&=&\displaystyle\frac{1}{\gamma_i}\ln \left( 1+\frac{\gamma_i}{(k+1-\frac{1}{N})\beta_i}\right)+
\frac{\gamma_i \sigma^2 (T-t_{n-2})}{2}(-2q_{n-2}^i+1)
\end{array}
\right.
\end{equation}
and Dealer $i$'s utility is given by
\begin{equation}\label{eq40}
\begin{array}{lll}
&&V^i\left(s_{n-2},x_{n-2}^i, \gamma_1,\cdots,\gamma_N,
q_{n-2}^1,\cdots,q_{n-2}^N, t_{n-2}\right)\\
&=&\displaystyle  -\exp(-\gamma_i(x_{n-2}^i+q_{n-2}^is_{n-2})) \exp\left(\frac{\gamma_i^2\sigma^2(q_{n-2}^i)^2(T-t_{n-2})}{2}\right)\\
&&\left[1-\frac{\gamma_i\Delta t_{n-2}}{(k+1-\frac{1}{N})\beta_i}\left(\lambda_{n-2}^{i,a}+
\lambda_{n-2}^{i,b}\right)
\right]h_{n-1}^i
\end{array}
\end{equation}
where $\Delta t_{n-2}=t_{n-1}-t_{n-2}$.
\end{proposition}

The proof can be found in Appendix C.

We remark that
\begin{description}
\item[(i)] The spread (compare to Eq. (\ref{eq35}))
$$
\delta_{n-2}^{i,b}+\delta_{n-2}^{i,a}
=\displaystyle\frac{2}{\gamma_i}\ln \left( 1+\frac{\gamma_i}{(k+1-\frac{1}{N})\beta_i}\right)+
\gamma_i \sigma^2 (T-t_{n-2})\\
$$
is independent of the inventory.
By taking a first-order approximation of the order arrival term, we have
(compare to Eq. (\ref{eq36}))
\begin{equation}\label{eq45}
\lambda_{n-2}^{i,b}+\lambda_{n-2}^{i,a}
=A\left[2-(k+1-\frac{1}{N})\beta_i(\delta_{n-2}^{i,a}+\delta_{n-2}^{i,b})-k\displaystyle\sum_{j\ne i}\beta_j(\delta_{n-2}^{j,a}+\delta_{n-2}^{j,b})+\cdots +\right].
\end{equation}
We notice that the linear term does not depend on the inventory.
Similar to the one-period case, substituting the linear approximation of $\lambda_{n-2}^{i,b}+\lambda_{n-2}^{i,a}$ into Eq. (\ref{eq39}),
one can get an approximation of Dealer $i$'s utility $f^i(s_{n-2},q_{n-2}^i,x_{n-2}^i,\gamma_1,\cdots,\gamma_N,t_{n-2})$,
which equals to
\begin{equation}\label{eq46}
\displaystyle  -\exp(-\gamma_i(x_{n-2}^i+q_{n-2}^is_{n-2})) \exp\left(\frac{\gamma_i^2\sigma^2(q_{n-2}^i)^2(T-t_{n-2})}{2}\right)
h_{n-2}^ih_{n-1}^i
\end{equation}
and it only depends on his own inventory.

\item [(ii)] Set
$$
f^i(s_{n-2},x_{n-2}^i,q_{n-2}^i,\gamma_1,\cdots,\gamma_N,t_{n-2})=-\exp\left(-\gamma_i(x_{n-2}^i+q_{n-2}^is_{n-2})\right)
g_{n-2}^i(q_{n-2}^i,t_{n-2})
$$
where
$$
g_{n-2}^i(q_{n-2}^i,t_{n-2})=\exp\left(\frac{\gamma_i^2\sigma^2(q_{n-2}^i)^2(T-t_{n-2})}{2}\right)h_{n-2}^ih_{n-1}^i
$$
is independent of stock price $s_{n-2}$ and cash wealth $x_{n-2}^i$.
\end{description}

By repeating the argument of this analysis,
one can get the following result for the multi-period model.

\subsubsection{The Multi-period Model}

Suppose that there are at most $N$ trades occur in $[t,T]$.
Divide the time period into $n+1$ small subintervals, $(t_0=t,t_1),\cdots,(t_{n-1},t_n),(t_n,T)$.
Assume that each trade may only occur immediately
after the beginning of those subintervals
and there is no trade occurring in $(t_n,T)$.
All dealers choose their bid and ask quotes at time $t_l$ $(l=0,1,\ldots,n-1)$,
defined by the controls $\delta_l^{i,b}$ and $ \delta_l^{i,a}$.
Adopting the approximation of dealers'  utility functions and using the back-forward analysis method, it is straightforward
to obtain the following proposition and we skip the proof.

\begin{proposition}
In the $n$-period model, dealers' optimal bid and ask quotes are given by
\begin{equation}\label{eq47}
\left\{
\begin{array}{lll}
\delta_l^{i,b}&=&\displaystyle\frac{1}{\gamma_i} \ln \left( 1+\frac{\gamma_i}{(k+1-\frac{1}{N})\beta_i}\right)+
\frac{\gamma_i \sigma^2 (T-t_l)}{2}(2q_l^i+1)\\
\delta_l^{i,a}&=&\displaystyle\frac{1}{\gamma_i} \ln \left( 1+\frac{\gamma_i}{(k+1-\frac{1}{N})\beta_i}\right)+
\frac{\gamma_i \sigma^2 (T-t_l)}{2}(-2q_l^i+1)
\end{array}
\right.
\end{equation}
and Dealer $i$'s utility is given by
\begin{equation}\label{eq48}
\begin{array}{lll}
&&V^i\left(s_l, x_l^i,\gamma_1,\cdots,\gamma_N,q_l^1,\cdots,q_l^N, t_{n-2}\right)\\
&=&\displaystyle  -\exp(-\gamma_i(x_l^i+q_l^is_l)) \exp\left(\frac{\gamma_i^2\sigma^2(q_l^i)^2(T-t_l)}{2}\right)\\
&&\left[1-\frac{\gamma_i\Delta t_l}{(k+1-\frac{1}{N})\beta_i}\left(\lambda_l^{i,a}+
\lambda_l^{i,b}\right)
\right]\prod_{m=l+1}^{n-1}h_m^i
\end{array}
\end{equation}
where $\Delta t_l=t_{l+1}-t_l$.
\end{proposition}

\subsubsection{The Continuous Model}

In every step of the back-forward model, we adopt the first-order approximation of the arrival terms appearing in the utility function.
Then we find that an approximate dealer's utility functions depend {\it only} on their own inventories.
We then consider the case of continuous model.
Define the approximate utility as $u_i(s,x_i,q_i,t)$.
The following theorem results from applying the principle of Dynamic Programming (DP).

\begin{theorem}
The optimal bid and ask quotes in dealer markets under competition are given by
\begin{equation}\label{eq49}
\left\{
\begin{array}{lll}
\delta_t^{i,b}&=&\displaystyle\frac{1}{\gamma_i}\ln \left( 1+\frac{\gamma_i}{(k+1-\frac{1}{N})\beta_i}\right)+
\frac{\gamma_i \sigma^2 (T-t)}{2}(2q_i+1)\\
\delta_t^{i,a}&=&\displaystyle\frac{1}{\gamma_i}\ln \left( 1+\frac{\gamma_i}{(k+1-\frac{1}{N})\beta_i}\right)+
\frac{\gamma_i \sigma^2 (T-t)}{2}(-2q_i+1)
\end{array}
\right.
\end{equation}
and dealers' approximate utility functions
under the quoting strategy are 
greater than those for the inactive case; that is,
\begin{equation}\label{eq50}
\begin{array}{lll}
u_i(s,x_i,q_i,t)&>& \displaystyle  -\exp(-\gamma_i(x_i+q_is))
\exp\left(\frac{\gamma_i^2q_i^2\sigma^2(T-t)}{2}\right).
\end{array}
\end{equation}
\end{theorem}

\begin{proof}
We note that
$$
\begin{array}{lll}
u_i(s,x_i,q_i,t)&=& \displaystyle \max_{\delta_i^a,\delta_i^b} E_t[-\exp(-\gamma_i(X_T^i+q_T^iS_T))]
\end{array}
$$
which is derived when the dealer follows the optimal strategy for setting $\delta_i^b$ and $\delta_i^a$ at each point in time period $[t,T]$.
To simplify our discussion,
we suppose that other dealers are in equilibrium under
the Cournot competitive environment \cite{Holt}.
The Cournot competition model is an economic setting for describing
a market where firms compete on their amount of output and make decisions independently of each other.

Using the principle of Dynamic Programming (DP) and under certain smoothness conditions of the value function $u_i$,the following HJB equation is obtained.
\begin{equation}\label{eqHJB51}
\left \{
\begin{array}{l}
\displaystyle \frac{\partial u_i}{\partial t} + \frac{1}{2}\sigma^2 \frac{\partial^2 u_i}{\partial s^2} + \max_{\delta_i^b}
\lambda_i^b [u_i(s, x_i-s+\delta_i^b, q_i+1,t)-u_i(s, x_i,q_i,t)]\\
 +\displaystyle  \max_{\delta_i^a} \lambda_i^a [u_i(s, x_i + s+\delta_i^a, q_i-1,t)
 -u_i(s, x_i,q_i,t)]=0\\
u_i ( s , x_i, q_i , T)=-\exp(-\gamma_i (  x_i +q_is)).
\end{array}
\right.
\end{equation}
As in \cite{Avellaneda08}, the following ansatz is considered 
$$
u_i(s,x_i,q_i,t) = -\exp(-\gamma_i x_i)\exp ( -\gamma_i \theta^i(s,q_i,t))
$$
where $\theta^i(s,q_i,t)$ is an approximate quadratic polynomial in the inventory variable $q_i$. Then the HJB equation can be written as follows:
\begin{equation}\label{eqHJB52}
\left \{
\begin{array}{l}
\displaystyle\theta_t^i + \frac{1}{2}\sigma^2\theta_{ss}^i - \frac{\sigma^2 \gamma_i (\theta_s^i)^2}{2}
+ \max_{\delta_i^b}\left\{  \frac{\lambda_i^b(\delta_1^b,\cdots,\delta_N^b)}{\gamma_i}\left(1-\exp(\gamma_i(s-\delta_i^b-r_i^b)\right) \right\}\\
+\displaystyle \max_{\delta_i^a}\left\{  \frac{\lambda_i^a(\delta_1^a,\cdots,\delta_N^a)}{\gamma_i}\Big(1-\exp(-\gamma_i(s+\delta_i^a-r_i^a)\Big) \right\}=0\\
\theta^i(s,q_i,T)=q_is.
\end{array}
\right.
\end{equation}
From the first-order optimality condition in Eq. (\ref{eqHJB52}),
we can obtain the optimal distance $\delta_i^a$ and $\delta_i^b$ of Dealer $i$
given the equilibrium values of all dealers, which satisfy
\begin{equation}\label{eqOP53}
\left \{
\begin{array}{l}
\displaystyle s-r_i^b(s,q_i,t)= \delta_i^b-\frac{1}{\gamma_i}\ln\left(1-\frac{\gamma_i\lambda_i^b}
{\frac{\partial}{\partial \delta_i^b}\lambda_i^b(\delta_1^b,\cdots,\delta_N^b)}\right)\\
\displaystyle r_i^a(s,q_i,t) - s= \delta_i^a-\frac{1}{\gamma_i}\ln\left(1-\frac{\gamma_i\lambda_i^a}
{\frac{\partial}{\partial \delta^a_i}\lambda_i^a(\delta_1^a,\cdots,\delta_N^a)}\right)\\
\end{array}
\right.
\end{equation}
where
\begin{equation}\label{eq54}
\left\{
\begin{array}{l}
r_i^b(s,q_i,t)=\theta^i(s,q_i,t)-\theta^i(s,q_i-1,t)\\
r_i^a(s,q_i,t)=\theta^i(s,q_i+1,t)-\theta^i(s,q_i,t).
\end{array}
\right.
\end{equation}
The above is the best response function of Dealer $i$
given the values of other dealers' quotes.
In Nash equilibrium, all dealers will play the best responses.
Thus we can solve the above $N$ equations simultaneously
to obtain the optimal feedback controls
$\delta_1^a,\cdots,\delta_N^a$ and $\delta_1^b,\cdots,\delta_N^b$.
We recall that
$$
\displaystyle \lambda_i^a= Ae^{-k(\beta_1\delta_1^a+\beta_2\delta_2^a+\cdots+\beta_N\delta_N^a)}e^{-\beta_i(1-\frac{1}{N})\delta_i^a}.
$$
Thus we have
\begin{equation}\label{eq55}
\displaystyle \frac{\partial \lambda_i^a}{\partial \delta_i^a}=-\left(k+1-\frac{1}{N}\right)\beta_i\lambda_i^a
\end{equation}
and similarly,
\begin{equation}\label{eq56}
\frac{\partial \lambda_i^b}{\partial \delta_i^b}=-\left(k+1-\frac{1}{N}\right)\beta_i\lambda_i^b.
\end{equation}
Substituting the optimal values given by Eq. (\ref{eqOP53}) into Eq.
(\ref{eqHJB52}) and using the rate of limit order book 
in Proposition $1$, we obtain
\begin{equation}\label{eqHJB57}
\left \{
\begin{array}{l}
\displaystyle\theta_t^i + \frac{1}{2}\sigma^2\theta_{ss}^i - \frac{\sigma^2 \gamma_i (\theta_s^i)^2}{2}
+\frac{\lambda_i^a+\lambda_i^b}{\left(k+1-\frac{1}{N}\right)\beta_i+\gamma_i} =0\\
\theta^i(s,q_i,T)=q_is.
\end{array}
\right.
\end{equation}


We consider an asymptotic expansion of
$\theta^i(s,q_i,t)$ in the inventory variable $q_i$
\begin{equation}\label{eq58}
\theta^i(s,q_i,t)=\theta^{i,(0)}(s,t)+\theta^{i,(1)}(s,t)q_i
+ \theta^{i,(2)}q_i^2+ \cdots +.
\end{equation}
From the exact relations of the indifference bid and ask prices,
$r_i^b$ and $r_i^a$, we obtain
\begin{equation}\label{eq59}
\left\{
\begin{array}{l}
r_i^a(s,q_i,t)=\theta^{i,(1)}(s,t)+(2q_i-1)\theta^{i,(2)}(s,t)+\ldots+\\
r_i^b(s,q_i,t)=\theta^{i,(1)}(s,t)+(2q_i+1)\theta^{i,(2)}(s,t)+\ldots+.
\end{array}
\right.
\end{equation}
Recall the first-order optimality conditions
\begin{equation}\label{eq60}
\left \{
\begin{array}{l}
\displaystyle s-r_i^b(s,q_i,t)= \delta_i^b-\frac{1}{\gamma_i}\ln\left(1+\frac{\gamma_i}{(k+1-\frac{1}{N})\beta_i}\right)\\
\displaystyle r_i^a(s,q_i,t) - s= \delta_i^a-\frac{1}{\gamma_i}\ln\left(1+\frac{\gamma_i}{(k+1-\frac{1}{N})\beta_i}\right)
\end{array}
\right.
\end{equation}
and we have
\begin{equation}\label{eq61}
\delta_i^b+\delta_i^a=-2\theta^{i,(2)}(s,t)+\frac{2}{\gamma_i}\ln\left(1+\frac{\gamma_i}{(k+1-\frac{1}{N})\beta_i}\right)
\end{equation}
which does not depend on the inventory $q_i$.
Taking a first-order approximation of the order arrival term
\begin{equation}\label{eq62}
\lambda_i^b+\lambda_i^a
=A\left[2-(k+1-\frac{1}{N})\beta_i(\delta_i^a+\delta_i^b)-k\displaystyle\sum_{j\ne i}\beta_j(\delta_j^a+\delta_j^b)+\cdots+\right]
\end{equation}
we notice that the linear term does not depend on  $q_i$.
Since
$$
\theta^i(s,0,t)=\theta^{i,(0)}
\quad {\rm and} \quad u_i(s,x_i,0,t)=g(x)
$$
then $\theta^{i,(0)}=0$.
Therefore, by grouping the terms of order $q_i$, we obtain
\begin{equation}\label{eq63}
\left\{
\begin{array}{l}
\theta_t^{i,(1)}+\frac{1}{2}\sigma^2\theta_{ss}^{i,(1)}=0\\
\theta^{i,(1)}(s,T)=s
\end{array}
\right.
\end{equation}
whose solution is $\theta^{i,(1)}(s,t)=s$. Grouping terms of order $q_i^2$ yields
\begin{equation}\label{eq64}
\left\{
\begin{array}{l}
\theta_t^{i,(2)}+\frac{1}{2}\sigma^2\theta_{ss}^{i,(2)}-\frac{1}{2}\sigma^2\gamma_i(\theta_s^{i,(1)})^2=0\\
\theta^{i,(2)}(s,T)=0\\
\end{array}
\right.
\end{equation}
whose solution is
\begin{equation}\label{eq65}
\theta^{i,(2)}(s,t)=-\frac{1}{2}\sigma^2\gamma_i(T-t).
\end{equation}
We obtain almost the same value function for an active agent as for an inactive agent,
$$
u_i(s,x_i,q_i,t)\approx -\exp(-\gamma_ix_i)\exp(-\gamma_iq_is)\exp\left(\frac{\gamma_i^2 q_i^2\sigma^2(T-t)}{2}\right)
$$
and the same indifference price
$$
r_i(s,q_i,t)=s-q_i\gamma_i\sigma^2(T-t).
$$

Now we analyze the difference between our approximation and the exact solution of our HJB equation.
Suppose that
\begin{equation}\label{eq66}
\theta^i(s,q_i,t)=w^{q_i}(t)+sq_i-\frac{1}{2}\sigma^2\gamma_i(T-t)q_i^2
\end{equation}
where $(w^{q_i})_{q_i\in \mathbf{N}}$ is a family of continuous functions.
Substituting the above expression directly into the HJB equation,
we obtain
\begin{equation}\label{eq67}
\left \{
\begin{array}{lll}
w_t^{q_i}+g^{q_i}(t)=0\\
w^{q_i}(T)=0
\end{array}
\right.
\end{equation}
where $(g^{q_i}(t))_{q_i\in \mathbf{N}}$ is a family of positive functions. Consequently,
$$
\begin{array}{lll}
w^{q_i}(t)= \displaystyle \int_t^Tg^{q_i}(u)du
\end{array}
$$
which is always greater than zero for $t<T$.
Thus we have
$$
\theta^i(s,q_i,t) > sq_i-\frac{1}{2}\sigma^2\gamma_i(T-t)q_i^2
$$
i.e.,
\begin{equation}\label{eq68}
\begin{array}{lll}
u_i(s,x_i,q_i,t)&>& \displaystyle  -\exp(-\gamma_i(x_i+q_is))\exp\left(\frac{\gamma_i^2q_i^2\sigma^2(T-t)}{2}\right).
\end{array}
\end{equation}

Now we set the bid-ask spread as
\begin{equation}\label{eq69}
\delta_i^a+\delta_i^b=\gamma_i\sigma^2(T-t)+\frac{2}{\gamma_i}\ln\left(1+\frac{\gamma_i}{(k+1-\frac{1}{N})\beta_i}\right)
\end{equation}
and a price adjustment as
\begin{equation}\label{eq70}
m_i  =  \delta_i^a-\delta_i^b = r_i^a+r_i^b-2s = 2r_i-2s
=-2q_i\gamma_i\sigma^2(T-t).
\end{equation}
\end{proof}

We remark that
\begin{description}
\item[(i)] For the ``frozen inventory'' problem, we have
$$
\left\{
\begin{array}{l}
E_t[x_i+q_iS_T]=x_i+q_is\\
u_i(s,x_i,q_i,t)=-\exp(-\gamma_ix_i)\exp(-\gamma_iq_is)\exp\left(\frac{\gamma_i^2q_i^2\sigma^2(T-t)}{2}\right).
\end{array}
\right.
$$
For the active dealer, we have
$$
\left\{
\begin{array}{lll}
E_t[X_i(T)+q_i(T)S_T]=x_i+q_is+E_t\left[\int_t^T\delta_i^adN_i^a +\int_t^T \delta_i^b dN_i^b \right]>x_i+q_is \\
u_i(s,x_i,q_i,t)>-\exp(-\gamma_ix_i)\exp(-\gamma_iq_is)\exp\left(\frac{\gamma_i^2q_i^2\sigma^2(T-t)}{2}\right).
\end{array}
\right.
$$
This means that active dealers using our strategy to quoting always
have an advantage over inactive dealers.

\item [(ii)] When $N=1$, then $\beta_i=1$,
$$
\delta_i^b+\delta_i^a=\gamma_i\sigma^2(T-t)+\frac{2}{\gamma_i}\ln\left(1+\frac{\gamma_i}{k}\right)
$$
which coincides with the results in
Avellanede and Stoikov (2008) \cite{Avellaneda08}.

\item[(iii)] The price adjustment $m_i=-2q_i\gamma_i\sigma^2(T-t)$,
depends on the dealer's inventory.
It is  an inventory response equation that specifies the price adjustments variable be negative (positive) when inventory is positive (negative).
When $m_i<0$, both bid price and ask price are ``low'', in this situation,
the dealer prefers to sell than to  purchase, and this will therefore reduce the dealer's inventory.
On the other hand, if $m_i>0$,
then the dealer prefer to purchase than to sell.
The degree of price response to an inventory change
depends on the same factors determining the size of the
bid and ask spread-time remained ($T-t$),
Dealer $i$'s risk aversion (determined by $\gamma_i$) and variance
(determined by $\sigma$);

\item[(iv)] Compared with the ``frozen inventory strategy'',
our strategy can eventually improve dealer's final
profit which is no less than the original indifference curve
(in the ``frozen inventory'' problem,
$(\delta_i^a, \delta_i^b)$ can be seen as $(+\infty,+\infty)$).
\end{description}

\section{Numerical Experiments}

In this section, we present and discuss numerical results on
both single monopolistic dealer and multiple dealers
in a competitive market.

\subsection{Bid-ask Quotes}

Avellaneda and Stoikov (2008) tested the performance of their ``inventory'' strategy, focusing primarily on the shape of the $P\&L$ profile and the final inventory $q_T$.
They compared it with a benchmark strategy that is symmetric around the
mid-price, regardless of the inventory under the assumption of a monopolistic dealer.
In this section, we test the performance of our strategy
for multiple dealers in a competitive stock market.

Suppose that there are $N$ dealers in a market.
In the numerical experiments, we assume $\beta_i$'s to be identical,
i.e., $\beta_i=1/N$.
As far as our simulation is concerned,
we chose the following parameters:
$s=100$, $t=0$, $T=1$, $\sigma=2$, $dt=0.005$, $q_i=0$, $\gamma_i=0.1$, $k=1.5$ and $A=140$, where $i=1,2,\ldots,N$
(the values of the parameters are chosen to be the same as those in \cite{Avellaneda08}).
The simulation results are obtained through the following procedures:

\begin{enumerate}
\item[(i)] At time $t$, the agents' quotes, $\delta_i^b$ and $\delta_i^a$ for Dealer $i$, $(i=1,2,\ldots,N)$ are computed, given the state variables;

\item[(ii)] At time $t+dt$, the state variables are updated: with probability $\lambda_i^a(\delta_i^a)dt$, Dealer $i$'s inventory decreases by one
and his wealth increases by $s+\delta_i^a$;
with probability $\lambda_i^b(\delta_i^b)dt$, Dealer $i$'s inventory variable increases by one and the wealth decreases by $s-\delta_i^b$,
where $i=1,2,\ldots,N$.
The mid-price is updated by a random increment $\pm\sigma\sqrt{dt}$.
\end{enumerate}

Figure 1 and Figure 2, respectively, present the optimal bid-ask quotes
and their profits of the monopolistic dealer and competitive dealers (two dealers).
The profit of each dealer in the two-dealer case is approximately
half of that in the monopolistic case.
This seems consistent with the assumption that $\beta_i=0.5$, ($i=1,2$),
equal sharing of the profits.
Avellaneda-Stoikov's inventory strategy generates persistent positive returns while our extended strategy preserves this good property
in a competitive market.

\begin{figure}[H]\label{fig1}
\centering
\includegraphics[width=7.0in,height=2.5in]{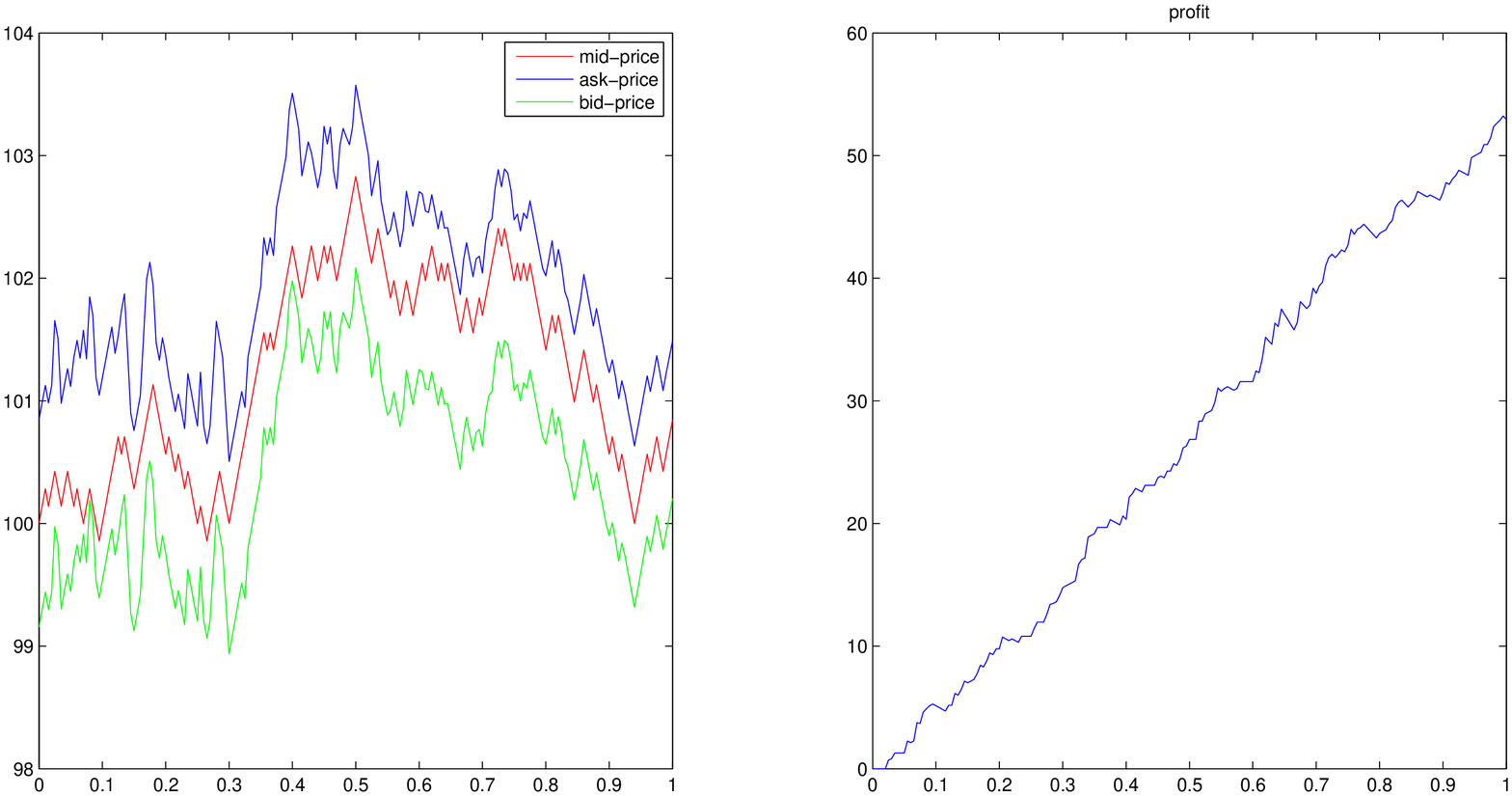}
\caption{The mid-price and the optimal bid-ask quotes of one monopolistic dealer.}
\end{figure}

\begin{figure}[H]\label{fig2}
\centering
\subfigure[Dealer 1]{
\includegraphics[width=7.0in,height=2.5in]{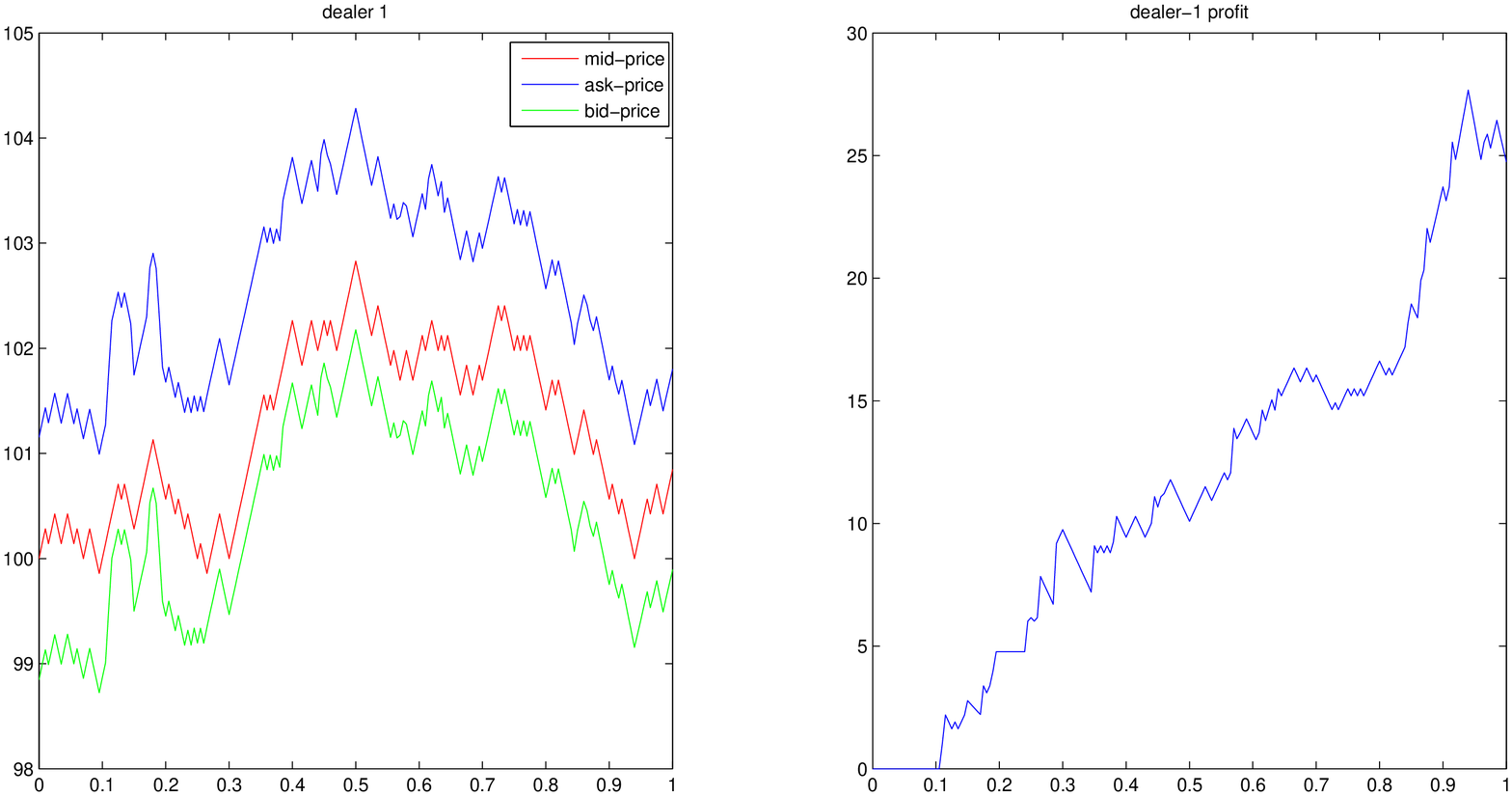}}
\subfigure[Dealer 2]{
\includegraphics[width=7.0in,height=2.5in]{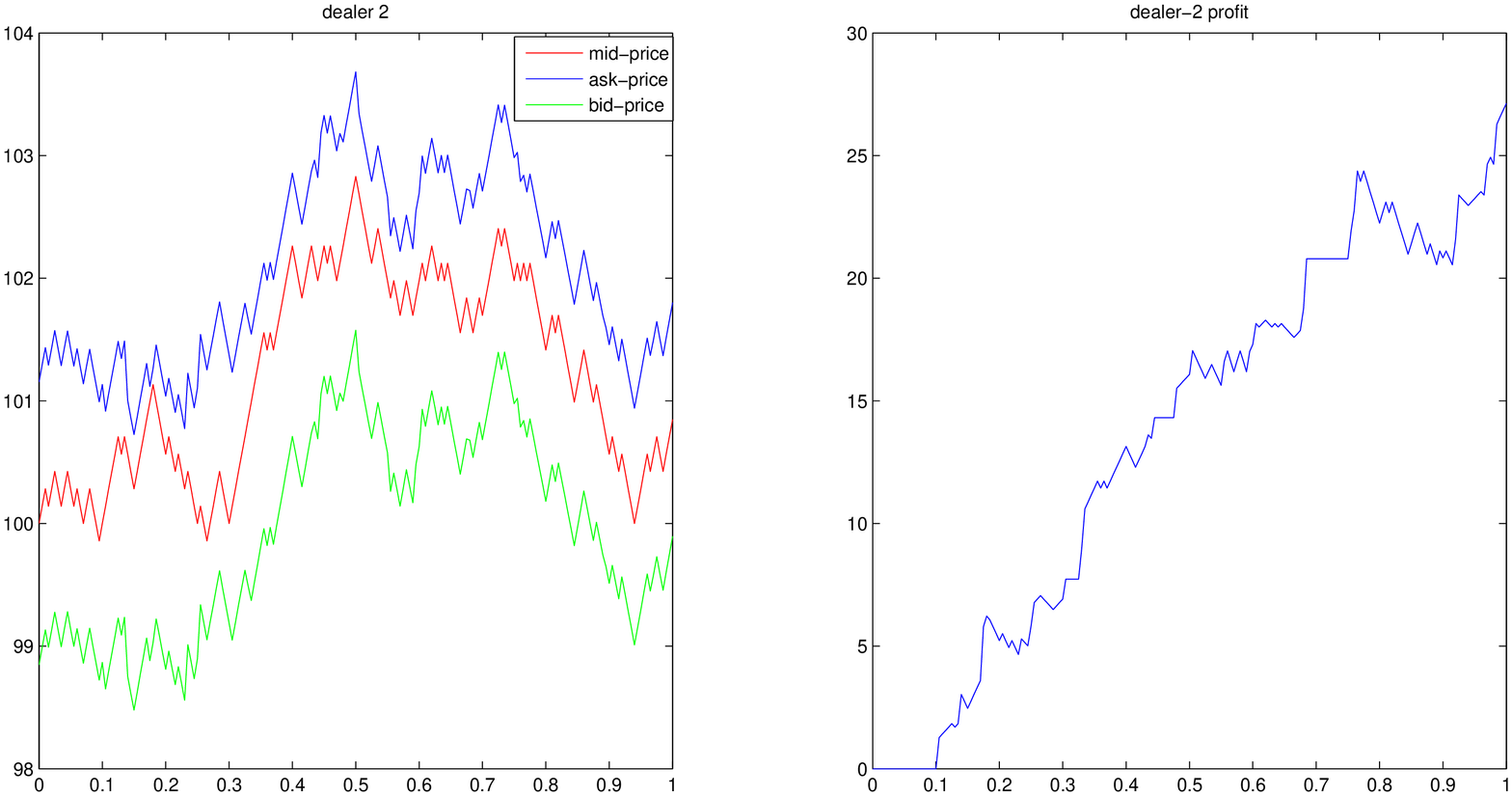}}
\caption{The mid-price and the optimal bid-ask quotes of two dealers.}
\end{figure}

\subsection{Competitive Dealers and Their Profits}

We consider the effects of changing from
one monopolistic dealer to multiple dealers in a competitive
market where the number of dealers varies.
The numerical results are presented in Tables 1-4 and
also Figures 3 and 4.

\begin{table}[htbp]
\caption{1000 simulations of one dealer with $\gamma=0.1$ and $\beta_1=1$.}
\centering
\begin{tabular}{lccccc}
\hline
Agent& Average Spread &Profit & Std (Profit) & $q_T$ &
Std ($q_T$)\\
\hline
Dealer &1.49&64.26&5.68&0.20&3.40\\
\hline
\end{tabular}
\end{table}

\begin{table}[htbp]
\caption{1000 simulations of two dealers with $\gamma_1=\gamma_2=0.1$ and $\beta_1=\beta_2=0.5$.}
\centering
\begin{tabular}{lccccc}
\hline
Agent& Average Spread& Profit & Std (Profit)& $q_T$ &Std ($q_T$)\\
\hline
Dealer 1&2.11&29.15&6.09&-0.02&2.88\\
Dealer 2&2.11&29.40&6.22&0.08&2.79\\
\hline
\end{tabular}
\end{table}

\begin{figure}[H]
\centering
\includegraphics[width=6.0in,height=3.in]{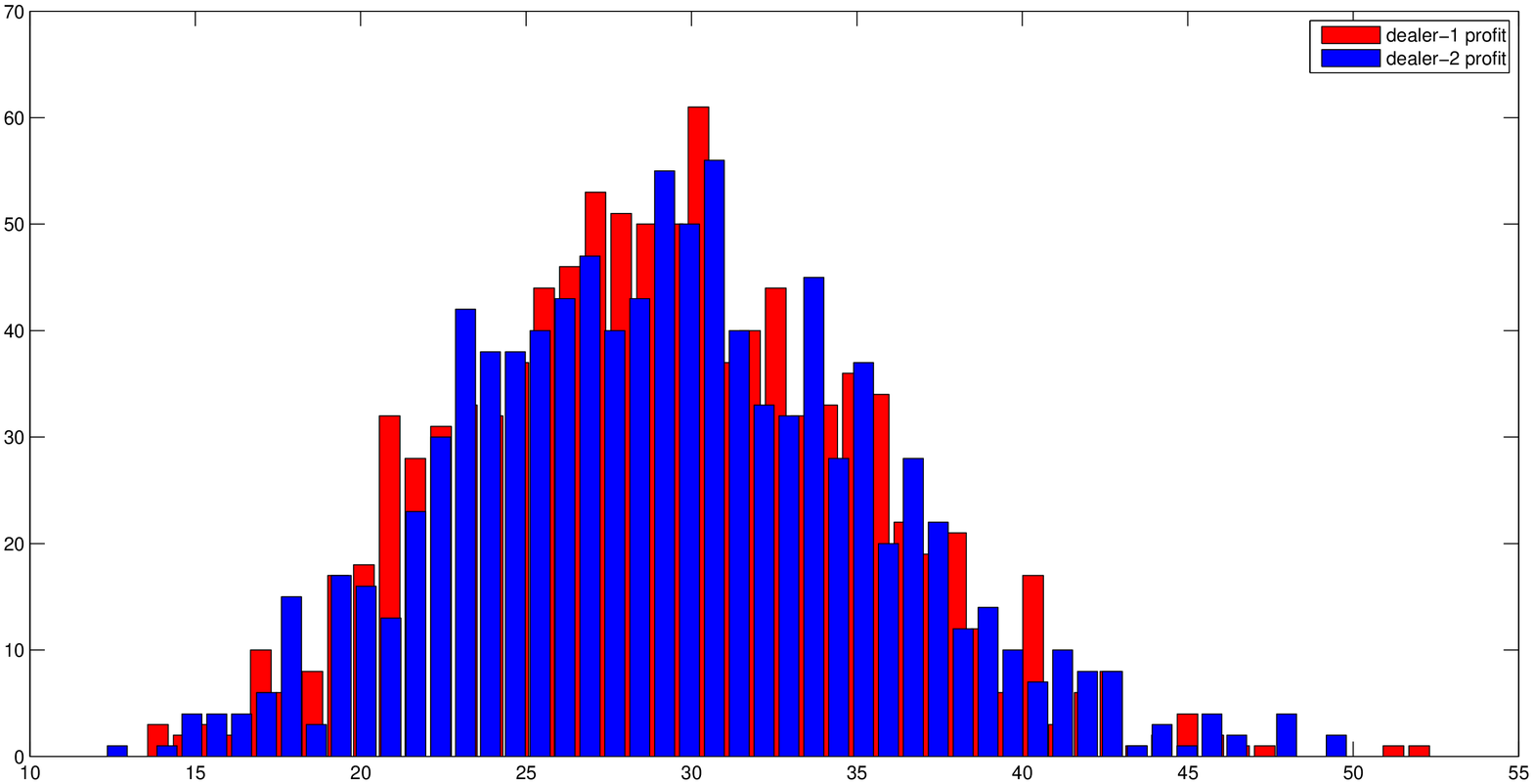}
\caption{ $\gamma_1=\gamma_2=0.1$ and $\beta_1=\beta_2=0.5$. }
\label{figure 3}
\end{figure}

\begin{table}[htbp]
\caption{ 1000 simulations of three dealers with $\gamma_1=\gamma_2=\gamma_3=0.1$ and $\beta_1=\beta_2=\beta_3=1/3$.}
\centering
\begin{tabular}{lccccc}
\hline
Agent & Average Spread & Profit & Std (Profit) & $q_T$ &Std ($q_T$)\\
\hline
Dealer 1&2.79&15.69&5.65&0.14&2.51\\
Dealer 2&2.79&15.85&5.69&-0.11&2.58\\
Dealer 3&2.79&15.88&5.53&-0.01&2.44\\
\hline
\end{tabular}
\end{table}

\begin{figure}[H]
\centering
\includegraphics[width=6.0in,height=3.in]{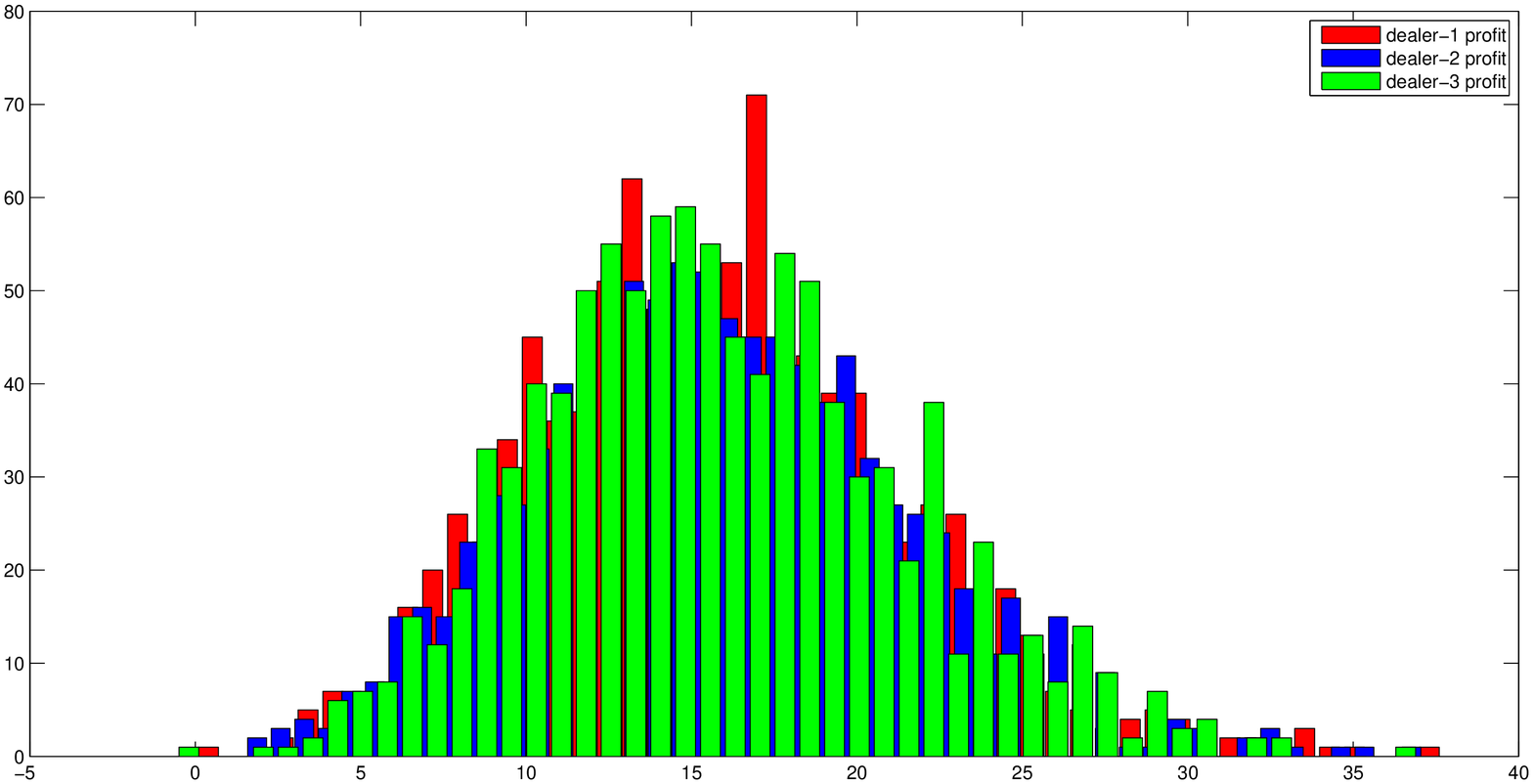}
\caption{ $\gamma_1=\gamma_2=\gamma_3=0.1$ and $\beta_1=\beta_2=\beta=1/3$. }
\label{figure 4}
\end{figure}

\begin{table}[htbp]
\caption{ 1000 simulations of seven dealers with $\gamma_i=0.1$ and $\beta_i=1/7$.}
\centering
\begin{tabular}{lccccc}
\hline
Agent& Average Spread & Profit & Std (Profit) &$q_T$ &Std ($q_T$)\\
 \hline
Dealer 1&5.40&1.76&2.34&0.028&0.79\\
Dealer 2&5.40&1.93&2.52&-0.045&0.84\\
Dealer 3&5.40&1.95&2.57&0.003&0.82\\
Dealer 4&5.40&1.79&2.46&-0.03&0.80\\
Dealer 5&5.40&1.83&2.49&0.003&0.79\\
Dealer 6&5.40&1.86&2.50&-0.001&0.83\\
Dealer 7&5.40&1.86&2.60&-0.019&0.86\\
\hline
\end{tabular}
\end{table}

We observe that the standard deviations of $q_T$ are relatively larger when
compared with the corresponding values of $|q_T|$ in most of the cases.
And in the case of seven dealers, the standard deviations of the profits are relatively large compared with the corresponding profits as well.
When the number dealers increases, we observe that profit decreases but the average spread increases.
Continuous markets are characterized by the bid and ask prices
at which trades can take place.
The bid-ask spread reflects the difference between
what active buyers must pay and what active sellers receive.
It is an indicator of the cost of trading and the illiquidity of a market.
We can see from the results that, when more and more agents
enter into a competitive market, on one hand,
competition becomes intense, which reduces the profit that each
agent can obtain from the stock market.
On the other hand, the liquidity of the stock enhances
which provides good supplies to the traders.

\subsection{Sensitivity Study of $\gamma$}

We now consider the effects of varying the parameter, $\gamma$.
Suppose the number of dealers in the competitive market is fixed and
that all dealers' initial states are the same, except the risk aversion parameter.
We can see from Tables 5 \& 6 and Figures 5 \& 6
that risky dealers take larger positions than risk-averse ones.
They have smaller values of average spreads and larger profits,
but also suffer from larger variances of profits and final inventories $q_T$, which lead to higher levels of uncertainty.

\begin{table}[H]
\caption{ 1000 simulations of two dealers with $\gamma_1=0.01$, $\gamma_2=1$ and $\beta_1=\beta_2=0.5$.}
\centering
\begin{tabular}{lccccc}
\hline
Agent& Average Spread& Profit &Std (Profit)& $q_T$ &Std ($q_T$)\\
\hline
Dealer 1&2.01&23.97&7.44&0.07&4.30\\
Dealer 2&3.39&17.98&4.25&-0.074&1.61\\
\hline
\end{tabular}
\end{table}

\begin{figure}[H]
\centering
\includegraphics[width=6.0in,height=3.in]{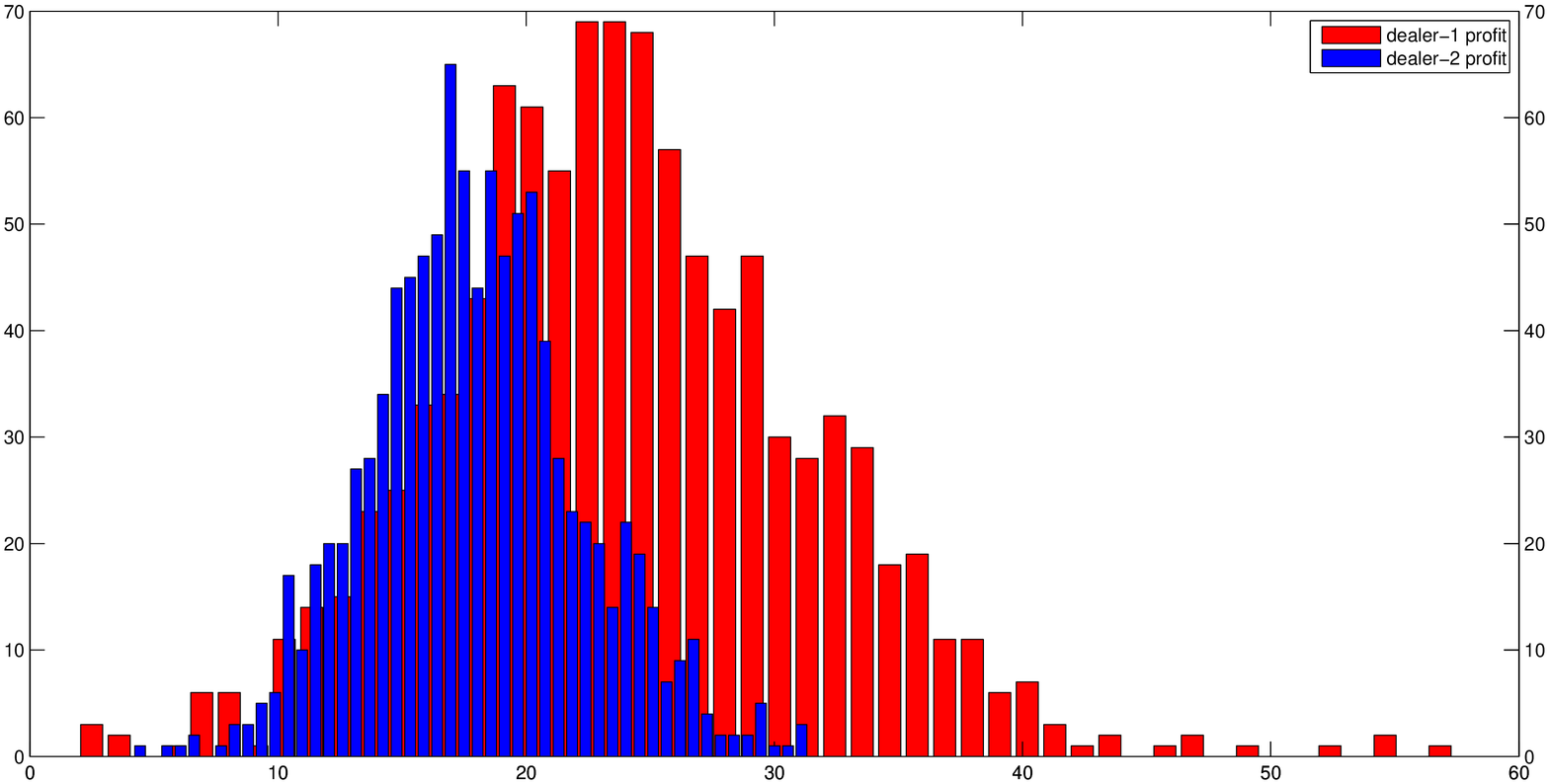}
\caption{ $\gamma_1=0.01$, $\gamma_2=1$ and $\beta_1=\beta_2=0.5$ }
\label{figure 5}
\end{figure}

\begin{table}[htbp]
\caption{ 1000 simulations of three dealers with $\gamma_1=0.01$, $\gamma_2=0.1$, $\gamma_3=1$ and $\beta_1=\beta_2=\beta_3=1/3$.}
\centering
\begin{tabular}{lccccc}
\hline
Agent& Average Spread& Profit& Std (Profit)&  $q_T$ &Std ($q_T$).\\
\hline
Dealer 1&2.77&14.36&6.26&-0.09&3.43\\
Dealer 2&2.79&14.21&5.68&-0.01&2.70\\
Dealer 3&3.74&10.99&3.74&0.08&1.47\\
\hline
\end{tabular}
\end{table}

\begin{figure}[H]
\centering
\includegraphics[width=6.0in,height=3.in]{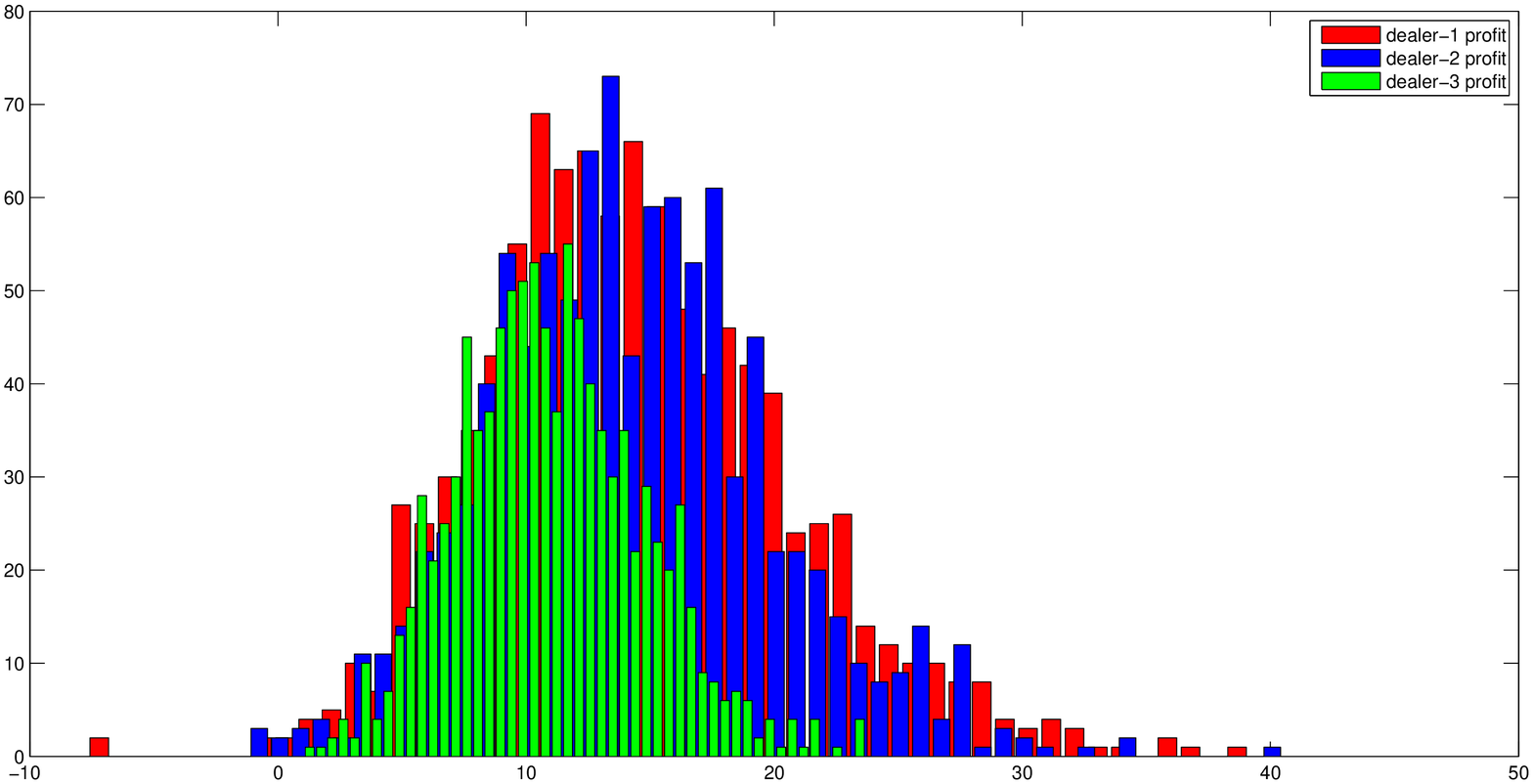}
\caption{ $\gamma_1=0.01$, $\gamma_2=0.1$, $\gamma_3=1$ and $\beta_1=\beta_2=\beta_3=1/3$ }
\label{figure 6}
\end{figure}

\subsection{Sensitivity Study of Initial Inventory Positions}

We also consider the effects of varying inventories on the performance of the dealers. For simplicity, we consider the two-dealer case.

\begin{table}[H]
\caption{1000 simulations of two dealers with $\gamma_1=\gamma_2=0.1$,
$\beta_1=\beta_2=0.5$, $q_1=10$ and $q_2=1$.}
\centering
\begin{tabular}{lccccc}
\hline
Agent & Average Spread & Profit &Std (Profit) & $q_T$ &Std ($q_T$)\\
\hline
Dealer 1&2.11&14.85&19.59&0.19&2.94\\
Dealer 2&2.11&30.24&6.34&-0.13&3.00\\
\hline
\end{tabular}
\end{table}

\begin{figure}[H]
\centering
\includegraphics[width=6.0in,height=3.in]{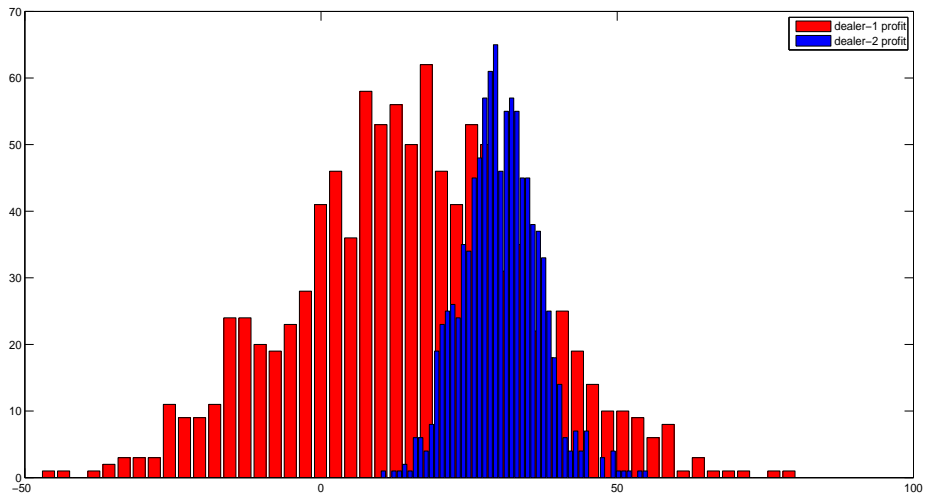}
\caption{ $\gamma_1=\gamma_2=0.1$, $\beta_1=\beta_2=0.5$, $q_1=10$ and $q_2=1$.}
\label{figure 7}
\end{figure}

\begin{table}[H]
\caption{ 1000 simulations of two dealers with $\gamma_1=\gamma_2=0.1$, $\beta_1=\beta_2=0.5$, $q_1=50$ and $q_2=0$.}
\centering
\begin{tabular}{lccccc}
\hline
Agent& Average Spread&Profit & Std (Profit)& $q_T$ & Std ($q_T$)\\
\hline
Dealer 1&2.11&-391.98& 84.14& 2.09&2.99\\
Dealer 2&2.11&37.55& 12.45&-1.98&2.91\\  \hline
\end{tabular}
\end{table}

\begin{table}[H]
\caption{ 1000 simulations of two dealers with $\gamma_1=\gamma_2=0.01$, $\beta_1=\beta_2=0.5$, $q_1=50$ and $q_2=0$.}
\centering
\begin{tabular}{lccccc}
\hline
Agent& Average Spread& Profit &Std (Profit)& $q_T$ & Std ($q_T$)\\
\hline
Dealer 1&2.01&13.11& 40.42& 25.31&4.74\\
Dealer 2&2.01&32.69& 16.18&-7.12&4.57\\
\hline
\end{tabular}
\end{table}

Setting all the parameters identical,
and we assume that dealers differ only in their initial inventory positions.
From Tables 7 \& 8 and Figure 7, one can see that with a larger initial position,
Dealer 1 in both examples gain smaller amounts of profits.
When the initial position is significantly large (Table 8),
Dealer 1 obtains a negative profit as one may prefer
a less risky position with a negative profit than
a risky position with a positive cash flow.
From Table 9, one can see that when lowering the risk aversion of Dealer 1,
his final profit will increase accordingly.
We refer interested readers on the analysis of the effects of initial inventories to the paper by Ho and Stoll (1980) \cite{Ho80}.

\section{Conclusions}

The field of market micro-structure encompasses two general types of models, i.e., inventory models and information models.
In this paper, we focus on the effect of inventory and extend Avellanede-Stoikov's optimal price strategy for
a monopolistic dealer to that for multiple dealers in a competitive market.
We derive the approximate optimal bid and ask prices for each dealer when the dealers are informed the severity of the competition.
We also analyze the effect of various parameters in our model on the bid-ask quotes and profits of the dealers in a competitive market.
For future research, one may take into account the presence of additional market factors such as order handling costs, asymmetric information and inter-dealer trading in the model.

In this paper, the mid-price of the stock is assumed to follow Eq. (\ref{eq1}).
In our future study, we shall consider a more general volatility model,
for example, the Heston stochastic volatility model \cite{Hull1,Hull2, Heston}.
We shall assume the mid-price follows the following
stochastic differential equations:
\begin{equation}\label{eq66}
\left\{
\begin{array}{lll}
dS_u = \sqrt{v_u} d W_u\\
dv_u=\theta (\alpha-v_u)du+\xi \sqrt{v_u}dB_u\\
dW_u\cdot dB_u=\rho du\\
\end{array}
\right.
\end{equation}
where $\theta$, $\alpha$ and $\xi$ are positive constants and
the volatility, $\sigma_u=\sqrt{v_u}$, is related to the level of stock price via the third equation in Eq. (\ref{eq66}) and $\rho$.

\section{Appendix}
\subsection{Appendix A.}
\begin{proof}
Consider first the arrival rate of market order.
When $N=1$,
$$
\lambda^a = \lambda^a(\delta^a)=A e^{-k\delta^a}.
$$
 By condition (i), when $N=2$
$$
\lambda^a=  \lambda^a(\delta_1^a,\delta_2^a)=f(\delta_1^a)^{\beta_1}f(\delta_2^a)^{\beta_2}
$$
where
$(\beta_1+\beta_2=1)$; when $\delta_1^a=\delta_2^a=\delta^a$, this is equivalent to the case when $N=1$, that is
$$
f(\delta^a)^{\beta_1}f(\delta^a)^{\beta_2}=\lambda^a(\delta^a,\delta^a)=\lambda^a(\delta^a)=A e^{-k \delta^a}
$$
Consequently
$$
f(\delta^a)=A e^{-k \delta^a}
\quad {\rm and} \quad
\lambda^a=\lambda^a(\delta_1^a,\delta_2^a)=A e^{-k(\beta_1 \delta_1^a+\beta_2 \delta_2^a)}.
$$
By Condition (i), for any $N$,
$$
\lambda^a=\lambda^a(\delta_1^a,\cdots,\delta_n^a)=f(\delta_1^a)^{\beta_1} \cdots f(\delta_N^a)^{\beta_N}
$$
where $\beta_1+\cdots+\beta_N=1$.
When  $\delta_1^a=\cdots=\delta_N^a=\delta^a$
then this situation is equivalent to the case when $N=1$, i.e.,
$$
\lambda^a(\delta^a,\cdots, \delta^a)=f(\delta^a)^{\beta_1} \cdots f(\delta^a)^{\beta_n}=\lambda^a(\delta^a)=A e^{-k \delta^a}
$$
and $f(\delta^a)=A e^{-k \delta^a}$. Consequently, we gave
$$
\lambda^a=\lambda^a(\delta_1^a,\ldots,\delta_N^a)
=A e^{-k(\beta_1 \delta_1^a +\cdots +\beta_N\delta_N^a)}.
$$
To calculate the arrival rate of buy and sell orders that will reach the dealer, we need the following information: \\
(i) the overall frequency of market orders;\\
(ii) the distribution of market orders' size;\\
(iii) the temporary impact of a large market order.\\
From above, we are given one of the estimation of market orders' frequency.
For the other conditions, from a lot of studies, see for instance
\cite{Gabaix06,Gopikrishnan00,Maslow01},
we have some statistical properties of the limit order book, such as,
the distribution of the size of market orders $Q$ obeys a power law:
$$
f^Q(x)\varpropto x^{-1-\alpha}
$$
and the market impact follows a ``$\log$ law'' \cite{Bouchaud02},
i.e., $\Delta p \varpropto \ln(Q)$ or
$$
\Delta p = \frac{1}{K_i} \ln(Q).
$$
Here we recall that
$\Delta p = p^Q -s$ where $p^Q$ is the price of the highest limit
order executed in the trade and $s$ is stock mid-price.
Aggregating the information of limit order book's statistical properties, we have
$$
\begin{array}{lll}
\lambda_i^a(\delta_i^a)&=&\lambda^a \cdot P(\Delta p>\delta_i^a)\\
&=&\lambda^a \cdot P(\ln(Q)>K_i\delta_i^a)\\
&=&\lambda^a \cdot p(Q>\exp(K_i\delta_i^a))\\
\end{array}
$$
and
$$
p(Q>\exp(K_i\delta_i^a)) \varpropto
 \int_{\exp{(K_i\delta_i^a)}}^{\infty} x^{-1-\alpha}dx.
$$
We have, for some constant $A$,
$$
\lambda_i^a(\delta_i^a)= A  \exp(-k(\beta_1\delta_1^a+\cdots+\beta_N\delta_N^a))
\cdot \exp(-(1-\frac{1}{N})\beta_i\delta_i^a).
$$
\end{proof}

\subsection{Appendix B}

\begin{proof}
We note that there is no trading in the period $(t_n,T)$,
$$
E\left[\exp(-\gamma_i\int_{t_n}^T (\delta_u^{i,a}dN_u^{i,a}+\delta_u^{i,b}dN_u^{i,b}))\Big|\mathcal{F}_{n-1}\right]=1.
$$
Thus the expected exponential utility of Dealer $i$'s ultimate wealth equals
$$
\begin{array}{lll}
E\left[-\exp(-\gamma_i(X_T^i+q_T^iS_T))| \mathcal{F}_{n-1}\right]\\
=-\Big\{ \lambda_{n-1}^{i,a}\Delta t_{n-1}\exp(-\gamma_i(x_{n-1}^i+q_{n-1}^is_{n-1}+\delta_{n-1}^{i,a}))\exp\left(\frac{\gamma_i^2\sigma^2(q_{n-1}^i-1)^2(T-t_{n-1})}{2}\right)\\
\lambda_{n-1}^{i,b}\Delta t_{n-1}\exp(-\gamma_i(x_{n-1}^i+q_{n-1}^is_{n-1}+\delta_{n-1}^{i,b}))\exp\left(\frac{\gamma_i^2\sigma^2(q_{n-1}^i+1)^2(T-t_{n-1})}{2}\right)\\
\left[1-\lambda_{n-1}^{i,a}\Delta t_{n-1} -\lambda_{n-1}^{i,b}\Delta t_{n-1}\right]\exp(-\gamma_i(x_{n-1}^i+q_{n-1}^is_{n-1}))\exp\left(\frac{\gamma_i^2\sigma^2(q_{n-1}^i)^2(T-t_{n-1})}{2}\right)\Big\}.
\end{array}
$$
By considering the first-order optimality conditions,
we can obtain the optimal bid and ask quotes as follows:
\begin{equation}\label{eq34}
\left\{
\begin{array}{lll}
\delta_{n-1}^{i,b}&=&\displaystyle\frac{1}{\gamma_i}\ln \left( 1+\frac{\gamma_i}{(k+1-\frac{1}{N})\beta_i}\right)+
\frac{\gamma_i \sigma^2 (T-t_{n-1})}{2}(2q_{n-1}^i+1)\\
\delta_{n-1}^{i,a}&=&\displaystyle\frac{1}{\gamma_i}\ln \left( 1+\frac{\gamma_i}{(k+1-\frac{1}{N})\beta_i}\right)+
\frac{\gamma_i \sigma^2 (T-t_{n-1})}{2}(-2q_{n-1}^i+1).
\end{array}
\right.
\end{equation}
Substituting the optimal bid and ask quotes into the expected utility function,
one can get the utility for Dealer $i$.
\end{proof}

\subsection{Appendix C}

\begin{proof}
We apply the principle of Dynamic Programming (DP) to
Dealer $i$'s utility function, and obtain
\begin{equation}\label{eq41}
\begin{array}{lll}
&&V^i\left(s_{n-2},\gamma_i,x_{n-2}^i, q_{n-2}^i,\gamma_j,q_{n-2}^j(j\ne i), t_{n-2}\right)\\
&=&\displaystyle \max_{\delta_{n-2}^{i,a},\delta_{n-2}^{i,b},\delta_{n-1}^{i,a},\delta_{n-1}^{i,b}} E\left[-\exp(-\gamma_i(X_T^i+q_T^iS_T))\Big|\mathcal{F}_{n-2}\right]\\
&=&\displaystyle \max_{\delta_{n-2}^{i,a},\delta_{n-2}^{i,b},\delta_{n-1}^{i,a},\delta_{n-1}^{i,b}} E\left[E[-\exp(-\gamma_i(X_T^i+q_T^iS_T))\big|\mathcal{F}_{n-1}]\Big|\mathcal{F}_{n-2}\right]\\
&=&\displaystyle \max_{\delta_{n-2}^{i,a},\delta_{n-2}^{i,b}}-\exp(-\gamma_i(x_{n-2}^i+q_{n-2}^is_{n-2}))
\Big\{\\
&&\lambda_{n-2}^{i,a}\Delta t_{n-2}\Big(\exp(-\gamma_i\delta_{n-2}^{i,a})\exp\left(\frac{\gamma_i^2\sigma^2
(q_{n-2}^i-1)^2\Delta t_{n-2}}{2}\right)g_{n-1}^i(q_{n-2}^i-1,t_{n-1})\Big)+ \\
&&\lambda_{n-2}^{i,b}\Delta t_{n-2}\Big(\exp(-\gamma_i\delta_{n-2}^{i,b})\exp\left(\frac{\gamma_i^2\sigma^2
(q_{n-2}^i+1)^2\Delta t_{n-2}}{2}\right)g_{n-1}^i(q_{n-2}^i+1,t_{n-1})\Big)+ \\
&&\left[1-\lambda_{n-2}^{i,a}\Delta t_{n-2}-\lambda_{n-2}^{i,b}\Delta t_{n-2}\right]\Big(\exp\left(\frac{\gamma_i^2\sigma^2
(q_{n-2}^i)^2\Delta t_{n-2}}{2}\right)g_{n-1}^i(q_{n-2}^i,t_{n-1})\Big)\Big\}\\
&=&\displaystyle \max_{\delta_{n-2}^{i,a},\delta_{n-2}^{i,b}}-\exp(-\gamma_i(x_{n-2}^i+q_{n-2}^is_{n-2}))
\exp\left(\frac{\gamma_i^2\sigma^2
(q_{n-2}^i)^2(T-t_{n-2})}{2}\right)h_{n-1}^i
\Big\{\\
&&\lambda_{n-2}^{i,a}\Delta t_{n-2}\exp(-\gamma_i\delta_{n-2}^{i,a})\exp\left(\frac{\gamma_i^2\sigma^2
(-2q_{n-2}^i+1)(T-t_{n-2})}{2}\right)+\\
&&\lambda_{n-2}^{i,b}\Delta t_{n-2}\exp(-\gamma_i\delta_{n-2}^{i,b})\exp\left(\frac{\gamma_i^2\sigma^2
(-2q_{n-2}^i+1)(T-t_{n-2})}{2}\right)+1-\lambda_{n-2}^{i,a}\Delta t_{n-2}-\lambda_{n-2}^{i,b}\Delta t_{n-2}\Big\}.
\end{array}
\end{equation}
By considering the first order optimality conditions,
we obtain Dealer $i$'s optimal ask quote as follows:
\begin{equation}\label{eq42}
\begin{array}{lll}
\delta_{n-2}^{i,a}&=&\displaystyle
\frac{1}{\gamma_i}\ln \left( 1+\frac{\gamma_i}{(k+1-\frac{1}{N})\beta_i}\right)+
\frac{\gamma_i \sigma^2 (T-t_{n-2})}{2}(-2q_{n-2}^i+1).
\end{array}
\end{equation}
Similarly, we can give his bid quote
\begin{equation}\label{eq43}
\begin{array}{lll}
\delta_{n-2}^{i,b}
&=&\displaystyle\frac{1}{\gamma_i}\ln \left( 1+\frac{\gamma_i}{(k+1-\frac{1}{N})\beta_i}\right)+
\frac{\gamma_i \sigma^2 (T-t_{n-2})}{2}(2q_{n-2}^i+1).
\end{array}
\end{equation}

Substituting the optimal bid and ask quotes into the utility function,
one can obtain Dealer $i$'s utility function:
\begin{equation}\label{eq44}
\begin{array}{lll}
&&V^i\left(s_{n-2},\gamma_i, x_{n-2}^i q_{n-2}^i,\gamma_j,q_{n-2}^j(j\ne i), t_{n-2}\right)\\
&=&\displaystyle  -\exp(-\gamma_i(x_{n-2}^i+q_{n-2}^is_{n-2})) \exp\left(\frac{\gamma_i^2\sigma^2(q_{n-2}^i)^2(T-t_{n-2})}{2}\right)\\
&&\left[1-\frac{\gamma_i\Delta t_{n-2}}{(k+1-\frac{1}{N})\beta_i}\left(\lambda_{n-2}^{i,a}+
\lambda_{n-2}^{i,b}\right) \right]h_{n-1}^i.
\end{array}
\end{equation}
\end{proof}

\section*{Acknowledgements}

The authors would like to thank the referees and the editor for
their helpful comments and suggestions.
This research work was supported by
Research Grants Council of Hong Kong under Grant
Number 17301214 and HKU CERG Grants
and Hung Hing Ying Physical Research Grant.


\begin{thebibliography}{1}
\bibitem{Avellaneda08}
M. Avellaneda and S. Stoikov (2008),
{\em High-frequency trading in a limit order book},
Quantitative Finance, 8, 217--224.

\bibitem{Bouchaud02}
J. Bouchaud, M. Mezard and M. Potters (2002),
{\em Statistical properties of stock order books: empirical results and
models}, Quantitative Finance, 2, 251--256.

\bibitem{CJ}
{\'A}. Cartea and S. Jaimungal (2015), 
{\em Risk measures and fine tuning of high frequency trading strategies}, Mathematical Finance, 25 576--611.

\bibitem{Copeland83}
T. Copeland and D. Galai (1983),
{\em Information effects on the bid-ask spread},
Journal of Finance, 38, 1457--1469.

\bibitem{Cohen81}
K. Cohen, S. Maier, R. Schwartz and D. Whitcomb (1981),
{\em Transaction costs, order placement strategy,
and existence of the bid-ask spread},
Journal of Political Economy, 89, 287--305.

\bibitem{Demsetz68}
H. Demsetz (1968),
{\em The cost of transacting},
Quarterly Journal of Economics, 82, 33--53.

\bibitem{Gabaix06}
X. Gabaix,  P. Gopikrishnan, V. Plerou and H. Stanley (2006),
{\em Institutional investors and stock market volatility},
Quarterly Journal of Economics, 121, 461--504.

\bibitem{Garman76}
M. Garman (1976),
{\em Market Microstructure},
Journal of Financial Economics, 3, 257--275.

\bibitem{Gopikrishnan00}
P. Gopikrishnan, V. Plerou, X. Gabaix, and H. Stanley (2000),
{\em Statistical properties of share volume traded in financial
markets}, Physical Review E, 62, 4493--4469.

\bibitem{GLF}
O. Gu{\'e}ant, C. Lehalle, and J. Fernandez-Tapia (2012),
 {\em Optimal portfolio liquidation with limit orders}, 
SIAM Journal on Financial Mathematics, 3, 740--764.

\bibitem{GP}
F. Guilbaud and H. Pham (2011), 
{\em Optimal high frequency trading with limit and market orders}, working paper.

\bibitem{Ho80}
T. Ho and H. Stoll (1980),
{\em On dealer markets under competition},
Journal of Finance, 35, 259--267.

\bibitem{Ho81}
T. Ho and H. Stoll  (1981),
{\em Optimal dealer pricing under transactions and return uncertainty},
Journal of Financial Economics, 9, 47--73.

\bibitem{Ho84}
T. Ho and R. Macris (1984),
{\em Dealer bid-ask quotes and transaction prices an empirical
study of some AMEX options},
Journal of Finance, 39, 23--45.

\bibitem{Holt}
C. Holt,
{\em Markets, games and strategic behavior}, Pearson Education, USA, 2006.

\bibitem{Hull1}
J. Hull and A. White (1990),
{\em Pricing interest-rate derivative securities},
The Review of Financial Studies, 3, 573--592.

\bibitem{Hull2}
J. Hull (2006),
{\em  Options, futures, and other derivatives},
N.J: Prentice Hall.  Upper Saddle River, (6th ed.).

\bibitem{Heston}
Heston, S.L.(1993),
{\em  A Closed-Form Solution for Options with Stochastic Volatility with Applications to Bond and Currency Options},
The Review of Financial Studies 6(2), 327--343.

\bibitem{Maslow01}
S. Maslow and M. Mills (2001),
{\rm Price fluctuations from the order
book perspective: empirical facts and a simple model},
Physica A,  299, 234--246.

\bibitem{Potters2003}
M. Potters and J. Bouchaud (2003),
{\em More statistical properties of order books and price impact},
Physica A, 324, 133--140.

\bibitem{Stoll78}
H. Stoll (1978)
{\em The supply of dealer services in securities markets},
Journal of Finance, 33, 1133--1151.

\bibitem{Tinic72}
S. Tinic (1972)
{\em The economics of liquidity services},
Quarterly Journal of Economics, 86, 79--93.

\bibitem{F.A2014}
F. Alavi Fard (2014)
{\em Optimal bid-ask spread in limit-order books under regime switching framework},
Review of Economics and Finance, 
Volume 4, Issue 4, Article ID: 1923-7529-2014-04-33-16.

\bibitem{N.Song2012}
N. Song, W. Ching, T. Siu and C. Yiu (2012)
{\em Optimal submission problem in a limit order book with VaR constraint},
The Fifth International Joint Conference on Computational Sciences and Optimization (CSO2012), IEEE Computer Society Proceedings, 266--270.

\end{thebibliography}
\end{document}